 \documentclass[trackchanges,twocolumn,times]{aastex7}

\expandafter\def\csname editcolor1\endcsname{magenta}% was magenta
\expandafter\def\csname editcolor2\endcsname{blue}  % was blue
\expandafter\def\csname editcolor3\endcsname{violet} % was violet

\usepackage{newtxtext,newtxmath}
\usepackage[T1]{fontenc}
\usepackage{graphicx}	% Including Figure\,files
\usepackage{amsmath}	% Advanced maths commands
\usepackage{float}
\usepackage{bookmark} %% generate table of contents sidebar in pdf 
\bookmarksetup{numbered, open,}
\usepackage{enumitem}
\setlist[enumerate]{itemsep=0mm}
\usepackage{hyperref}
\usepackage{subfigure}
\usepackage{xspace}
\usepackage{xcolor} 
\usepackage{soul} 
\usepackage{multirow}

\newcommand{\HeI}{He~{\sc i}}
\newcommand{\OI}{O~{\sc i}}
\newcommand{\OII}{O~{\sc ii}}

\newcommand{\CII}{C~{\sc ii}}
\newcommand{\CI}{C~{\sc i}}

\newcommand{\MgII}{Mg~{\sc ii}}

\newcommand{\SiI}{Si~{\sc i}}
\newcommand{\SiII}{Si~{\sc ii}}
\newcommand{\SiIII}{Si~{\sc iii}}

\newcommand{\SI}{S~{\sc i}}

\newcommand{\CaII}{Ca~{\sc ii}}

\newcommand{\FeI}{Fe~{\sc i}}
\newcommand{\FeII}{Fe~{\sc ii}}
\newcommand{\FeIII}{Fe~{\sc iii}}

\newcommand{\CoII}{Co~{\sc ii}}

\newcommand{\Cofs}{$^{56}$Co}
\newcommand{\Nifs}{$^{56}$Ni}

\newcommand{\Msun}{\ensuremath{\mathrm{M}_{\odot}}\xspace}

\newcommand{\kms}{km~s\ensuremath{^{-1}}\xspace}

\graphicspath{{./}{figs/}}

\usepackage{graphicx} 

\received{xxx}
\revised{xxx}
\accepted{xxx}

\submitjournal{AJ}

\shorttitle{The Hawaii Infrared Supernova Study}
\shortauthors{Medler, Ashall, et al.}

\begin{document}

\title{The Hawaii Infrared Supernova Study (HISS): Spectroscopic Data Release 1}

\correspondingauthor{Kyle Medler}
\email{kmedler@hawaii.edu}

\author[0000-0001-7186-105X]{K.~Medler}
\affiliation{Institute for Astronomy, University of Hawai'i at Manoa, 2680 Woodlawn Dr., Hawai'i, HI 96822, USA }
\email{kmedler@hawaii.edu}

\author[0000-0002-5221-7557]{C.~Ashall}
\affiliation{Institute for Astronomy, University of Hawai'i at Manoa, 2680 Woodlawn Dr., Hawai'i, HI 96822, USA }
\email{cashall@hawaii.edu}

\author[0000-0002-9301-5302]{M.~Shahbandeh}
\altaffiliation{STScI Fellow}
\affiliation{Space Telescope Science Institute, 3700 San Martin Drive, Baltimore, MD 21218-2410, USA}
\email{mshahbandeh@stsci.edu}

\author[0000-0002-7566-6080]{J.~M.~DerKacy}
\affiliation{Space Telescope Science Institute, 3700 San Martin Drive, Baltimore, MD 21218-2410, USA}
\email{jmderkacy@stsci.edu}

\author[0000-0003-3953-9532]{W.~B.~Hoogendam}
\altaffiliation{NSF Graduate Research Fellow}
\affiliation{Institute for Astronomy, University of Hawai'i at Manoa, 2680 Woodlawn Dr., Hawai'i, HI 96822, USA }
\email{willemh@hawaii.edu}

\author[0000-0002-6230-0151]{D.~O.~Jones}
\affiliation{Institute for Astronomy, University of Hawai'i, 640 N. A'ohoku Pl., Hilo, HI 96720, USA}
\email{dojones@hawaii.edu}

\author[0000-0003-4631-1149]{B.~J.~Shappee}
\affiliation{Institute for Astronomy, University of Hawai'i at Manoa, 2680 Woodlawn Dr., Hawai'i, HI 96822, USA }
\email{shappee@hawaii.edu}

\author[0000-0001-9668-2920]{J.~T.~Hinkle}
\altaffiliation{FINESST FI}
\affiliation{Institute for Astronomy, University of Hawai'i at Manoa, 2680 Woodlawn Dr., Hawai'i, HI 96822, USA }
\email{jhinkle6@hawaii.edu}

\author[0000-0002-7305-8321]{C.~M.~Pfeffer}
\altaffiliation{NSF Graduate Research Fellow}
\affiliation{Institute for Astronomy, University of Hawai'i at Manoa, 2680 Woodlawn Dr., Hawai'i, HI 96822, USA }
\email{cmpfeffer@vt.edu}

\author[0000-0001-5393-1608]{E.~Baron}
\affiliation{Planetary Science Institute, 1700 East Fort Lowell Road, Suite 106,
 Tucson, AZ 85719-2395 USA}
\affiliation{Hamburger Sternwarte, Gojenbergsweg 112, D-21029 Hamburg, Germany}
\email{ebaron@psi.edu}

\author[0000-0002-4338-6586]{P.~Hoeflich}
\affiliation{Department of Physics, Florida State University, 77 Chieftan Way, Tallahassee, FL 32306, USA}
\email{phoeflich@fsu.edu}

\author[0000-0003-1039-2928]{E.~Hsiao}
\affiliation{Department of Physics, Florida State University, 77 Chieftan Way, Tallahassee, FL 32306, USA}
\email{ehsiao@fsu.edu}

\begin{abstract}

We present the first data release of the Hawaii Infrared Supernova Study (\textit{HISS}), consisting of a large sample of near-infrared (NIR) spectra, $0.7 - 2.5$~\micron, obtained with the Keck-II/NIRES and IRTF/SpeX spectrographs. This sample is comprised of 90 NIR spectra of 48 transient events, spanning from hours after explosion to $\geq + 350$~days. Acquired over three years (2021–2024), this data release includes 17 Type Ia SNe, 15 Type II SNe, 8 Stripped Envelope SNe, 6 interacting SNe, 1 TDE, and 1 SLSN‑I. These spectra were all systematically reduced using either the \textsc{Python}-based reduction code \textsc{Pypeit} or the \textsc{IDL}-based \textsc{Spextool} and constitute one of the largest NIR samples of transients available to the astrophysical community. We show the utility of NIR spectra and identify the key spectral features across multiple types of SNe.
We show how both early-time and nebular-phase NIR spectra can be used to investigate the physics of the explosion, and to reveal the properties of the progenitor.
With the addition of this dataset, the number of publicly available NIR spectra spanning multiple transient types has been substantially increased.
In its next phase, \textit{HISS} will leverage target-of-opportunity spectral observations and NIR imaging from telescopes on Maunakea.
Expanding the NIR dataset of SNe is vital to the transient community, particularly in light of the increasing emphasis on the infrared regime following the recent launch of the \textit{James Webb Space Telescope} and the forthcoming launch of the \textit{Nancy Grace Roman Space Telescope}.

\end{abstract}


\section{Introduction} 
\label{sec:intro}

As some of the most energetic events in the universe, supernovae (SNe) play a crucial role in shaping their local environments. They influence their surroundings by driving gas out of their host galaxies \citep{Lanfranchi_2024} and by enhancing the synthesis of heavy elements, including those formed through the r-process \citep[e.g.,][]{Matteucci_1986, Arnaud_1992, Tsujimoto_1995, Maoz_2017, Johnson_2019, Haynes_2019, Cote_2019}. Understanding how SNe explode and disperse heavy elements into the surrounding interstellar medium \citep{Arnaud_1992} is essential for determining how high- and low-mass stars end their lives, as well as how SNe contribute to the chemical enrichment of their host galaxies \citep{Timmes_1995, Kobayashi_2009, Berg_2019}.

SNe are generally divided into two main categories: Type Ia SNe (SNe\,Ia), which result from the thermonuclear explosion of at least one Carbon-Oxygen White Dwarf (WD) \citep{Hoyle_1960}, and core-collapse SNe (CC-SNe), which occur when the core of a massive star collapses under its own gravity and a shockwave is launched from the forming proto-neutron star that unbinds and ejects the remaining material \citep{Baade_1934, Woosley_2002}.

Several formation channels have been proposed to explain SNe\,Ia. The first is the single degenerate system where material is accreted onto the WD from a non-degenerate companion, such as a red-giant or helium-star \citep{Whelan_1973, Iben_1984, Nomoto_1984, Han_2006}. 
The second popular formation channel involves the merger of two WDs in a compact binary system \citep{Iben_1984, Webbink_1984, Dong_2015}. 
In both cases, the total WD mass involved in the explosion may be sub-Chandrasekhar ($\mathrm{M_{Ch}}$), near-$\mathrm{M_{Ch}}$, or super-$\mathrm{M_{Ch}}$ \citep{Hoeflich_1996, Sim_2010, Glasner_2018, Townsley_2019}. To date, determining which progenitor channel dominates the SNe\,Ia population has remained an open question in the community.

In contrast, CC-SNe, which originate from stars with zero-age main sequence masses $\mathrm{M_{ZAMS} \geq 8\,M_\odot}$, display a range of properties depending on the composition of the progenitor prior to explosion \citep{Lyman_2016, Anderson_2019, Bi_2024}. CC-SNe can be further classified into one of two groups; the H-rich Type II SN (SN\,II) or the stripped-envelope SN (SE-SN) \citep{Filippenko_1997}. Unlike SNe\,II which retain thick hydrogen envelopes, the progenitors of SE-SNe have undergone mass-loss episodes prior to explosion stripping them of their outer envelope. This degree to which the progenitor is stripped dictated the type of SE-SNe which include; the H/He-rich Type IIb SN (SN\,IIb), H-poor/He-poor Type Ib SN (SN\,Ib), and the H/He-poor Type Ic SN (SN\,Ic). The mass-loss mechanism responsible for the formation of SE-SNe progenitors is thought to occur via stellar eruptions \citep{Smith_2006a, Smith_2006b}, metallicity dependent winds \citep{Heger_2003, Pauldrach_2012, Yoon_2015}, or mass transfer through binary interaction \citep{Podsiadlowski_1993, Pols_2002, Yoon_2015}.

Within CC-SNe, a small fraction of objects \citep{Frohmaier_2021} are observed to possess peak absolute magnitudes greater than $\mathrm{M_{peak}} \geq -21$~mag. These SNe are classified as a superluminous supernova (SLSN) \citep{Gal-Yam_2012}. The hydrogen-poor SLSNe (SLSN-I) are characterized by a blue continua and strong ``W''-shaped \OII\ lines \citep{Quimby_2011, Nicholl_2015, Gal-Yam_2019, Konyves-Toth_2021, Saito_2024}. The origins of these luminous objects are still heavily debated, with formation channels ranging from the pair-instability SN \citep{Barkat_1967, Chatzopoulos_2012}, the magnetar-powered SN \citep{Kasen_2010, Woosley_2010} or through the interaction between SN ejecta and a dense circumstellar medium (CSM) \citep{Chevalier_2011, Chatzopoulos_2013}.

However, CSM interaction is not limited to SLSNe. Prior to exploding, the progenitors of typical SNe can undergo significant mass-loss episodes, resulting in the buildup of a dense shell of CSM surrounding the progenitor \citep{Smith_2014}. After exploding and as the expanding ejecta collides with the surrounding CSM, the kinetic energy of the ejecta is converted into X-ray and ultraviolet radiation. These high-energy photons escape into the CSM, ionizing the dense material, forming narrow emission lines \citep{Chevalier_2003, Chandra_2018}. The narrow lines observed in the spectra of interacting SNe exhibit line velocities on the order of $\sim 100$~\kms\ \citep{Kiewe_2012, Taddia_2013, Moriya_2014} which is associated with the velocity of the CSM, much slower than the expansion velocities of the SN ejecta, $\sim 1,000 - 20,000$~\kms\ \citep{Takats_2012}. Signatures of this type of interaction has been observed in all types of SNe, including SNe\,Ia \citep{Aldering_2006, Silverman_2013}, SNe\,II \citep{Schegel_1990, Filippenko_1997, Chugai_2004, Taddia_2013, Moriya_2020}, and SE-SNe \citep{Pastorello_2007, Pellegrino_2022, Schulze_2024}, making them ideal for studying mass-loss in both high- and low-mass stars.

\begin{figure*}
\begin{minipage}[c]{0.5\linewidth}
\includegraphics[width=\linewidth]{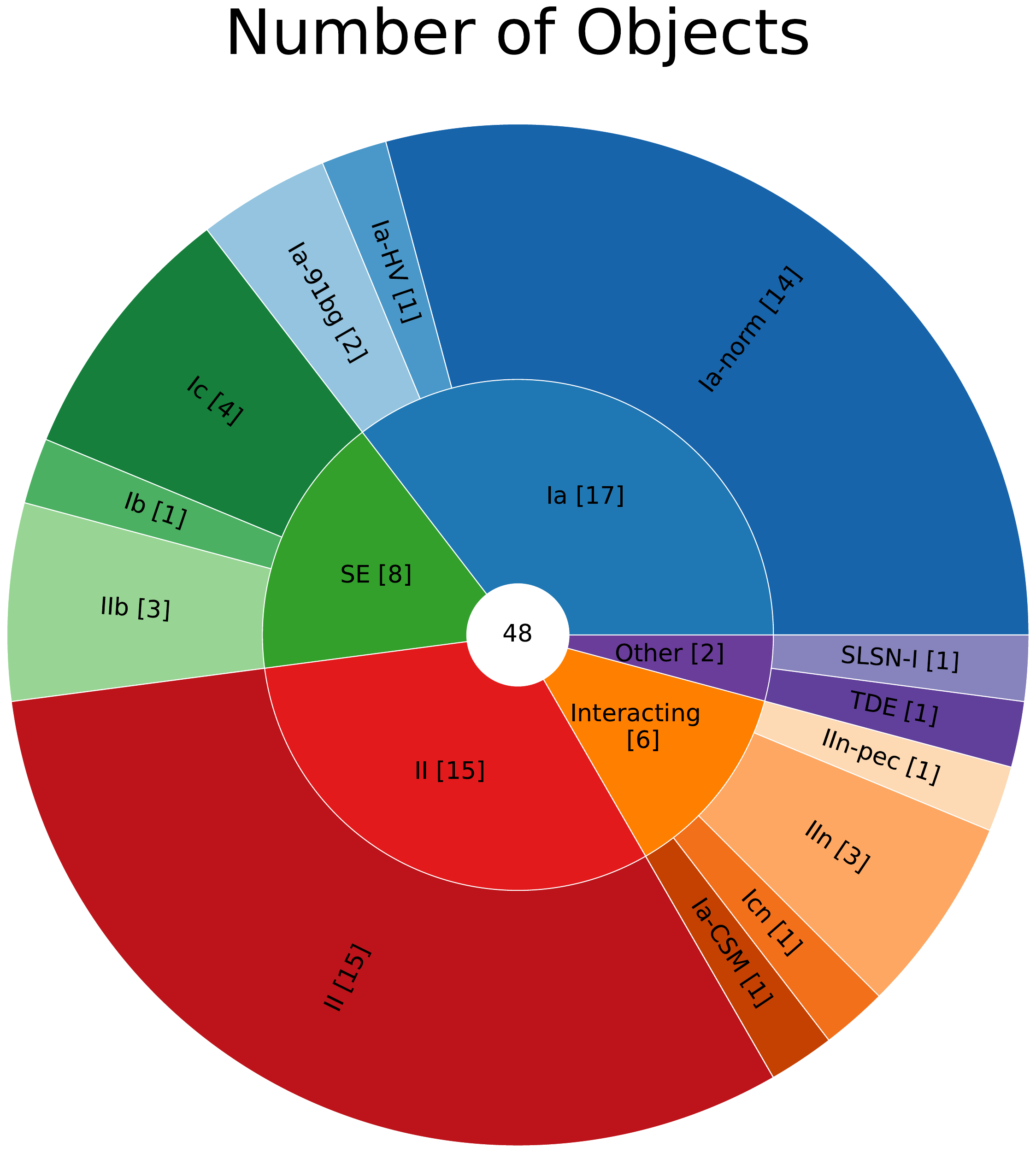}
\end{minipage}
\hfill
\begin{minipage}[c]{0.5\linewidth}
\includegraphics[width=\linewidth]{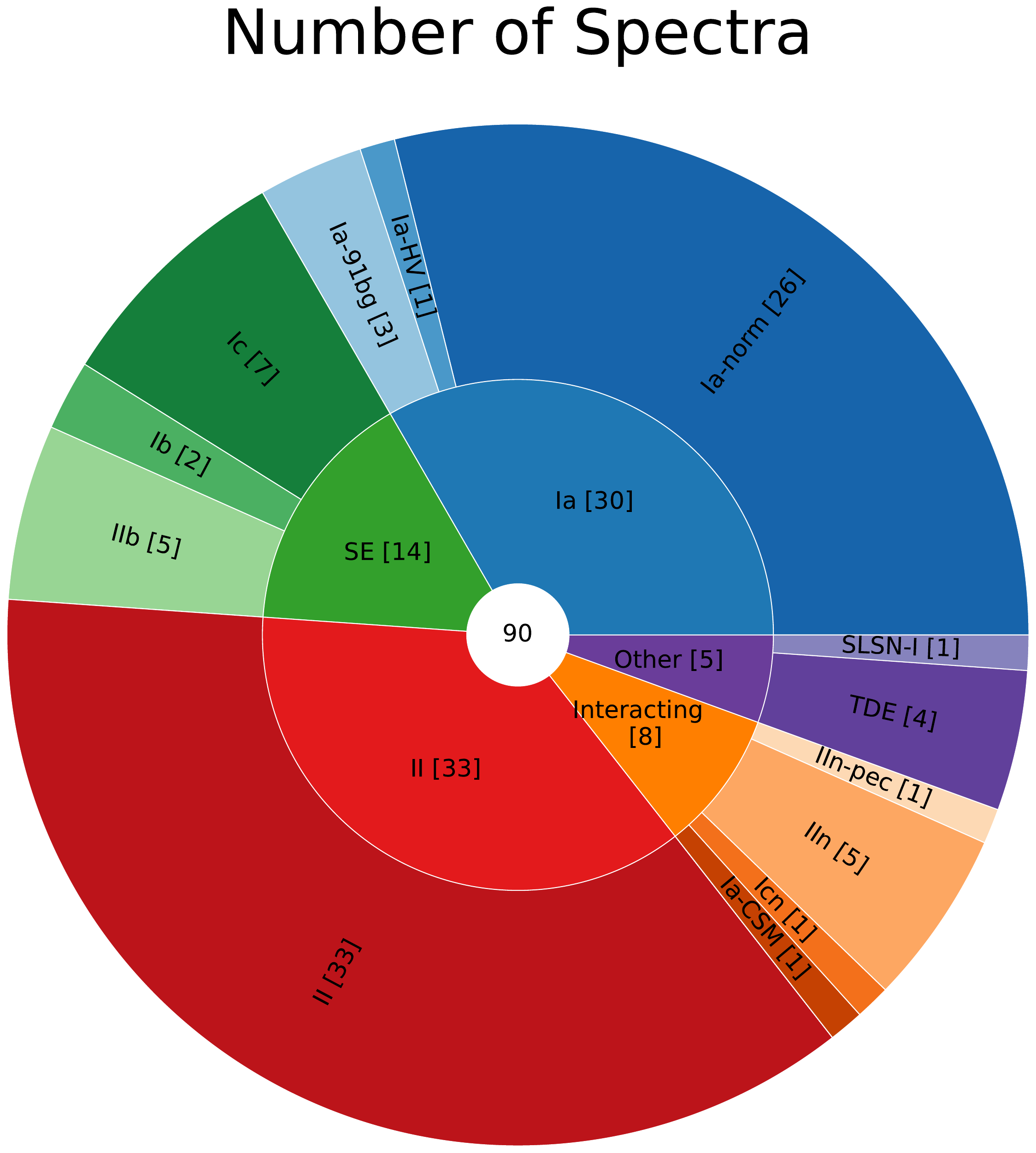}
\end{minipage}%
\caption{Left: The total number of objects observed in DR1 categorized by supernova types and sub-types. Right: The number of spectra obtained for each spectral classification and subtype.}
\label{fig:Pie_charts}
\makeatletter
\newcommand{\manuallabel}[2]{\def\@currentlabel{#2}\label{#1}}
\makeatother
\manuallabel{fig:Pie_chart}{1}
\end{figure*}

Traditionally, since SNe expel most of their radiative luminosity in the visible wavelength regime, they have been primarily detected and followed up in the optical \citep{Jha_2019}. This has been enabled by several large, high-cadence surveys such as the All-Sky Automated Survey for Supernovae \citep[ASAS-SN;][]{Shappee_2014}, the Asteroid Terrestrial-impact Last Alert System \citep[ATLAS;][]{Tonry_2018}, Panoramic Survey Telescope and Rapid Response System \citep[Pan-STARRS;][]{Chambers_2016}, and the Zwicky Transient Facility \citep[ZTF;][]{Bellm_2019}. Dedicated SNe classification and follow-up programs, including the Public ESO Spectroscopic Survey of Transient Objects \citep[PESSTO;][]{Smartt_2015}, the Carnegie Supernova Project \citep[CSP;][]{Hamuy_2006}, and the Spectroscopic Classification of Astronomical Transients survey \citep[SCAT;][]{Tucker_2022} have also played a crucial role in characterizing these events.

However, this emphasis on optical wavelengths omits crucial information in other regimes, such as the ultraviolet \citep[e.g.,][]{Panagia_2007, Bufano_2009, Ashall18, DerKacy_2023b, Hoogendam_2024} and the near-infrared (NIR).
In particular, NIR observations of SNe reveal several spectral lines associated with elements that lack optical features or whose optical lines are heavily blended at shorter wavelengths \citep{Marion_2009, Davis_2019, Hsiao_2019, Shahbandeh_2022, Lu_2023, DerKacy_2023a, DerKacy_2024, Tinyanont_2024, Ashall_2024, Hoogendam_2025a, Hoogendam_2025b, Kumar25}. These features contain vital clues on the physics of SN explosions, and the nature of their progenitor systems \citep{Maeda_2010, Hachinger_2012, Dessart_2020, Williamson_2021, Shahbandeh_2022, Dessart_2024}.

Several previous works have obtained large ground-based NIR datasets of SNe.
These include: \textit{i)} \citet{Marion_2009}, which published a sample of 41 spectra obtained from 28 normal SNe\,Ia, \textit{ii)} the Carnegie Supernova Project II \citep[CSP II;][]{Phillips_2019, Hsiao_2019, Davis_2019, Ashall21, Hoogendam_2022, Shahbandeh_2022, Lu_2023}  which obtained 909 spectra of 249 thermonuclear and core-collapse SNe between 2011 and 2015, and \textit{iii)} the Keck Infrared Transient Survey \citep[KITS;][]{Tinyanont_2024} which obtained 274 spectra of 146 transients ranging from all types of thermonuclear SNe, to CC-SNe, several tidal disruption events (TDEs)\setcounter{footnote}{0}\footnote{A TDE occurs when a star passes within the tidal radius of a supermassive black hole. The star is torn apart as the tidal forces overwhelm the star's self-gravity and the fallback of the disrupted material onto the black hole powers a luminous accretion flare \citep[e.g.,][]{Rees_1988, Evans_1989}.} and a luminous red nova.

Despite previous ground-based NIR follow-up campaigns, there is still a dearth of NIR spectra relative to optical observations of transients. 
The launch of the \textit{James Webb Space Telescope (JWST)} and the upcoming launch of the \textit{Nancy Grace Roman Space Telescope (Roman)} will allow us to study fainter and older transients than previously possible. However, achieving the full capabilities of these flagship missions requires a robust understanding of the NIR behavior of transients through large reference samples obtained with ground-based telescopes.

Here, we present the first data release (DR1) from the \textit{Hawaii Infrared Supernova Study (HISS)}, which utilizes the NIR capabilities of Maunakea telescopes to assemble a large sample of SNe spanning all types. This first data release covers the period from 2021 to 2024. The goal of the first stage of \textit{HISS} was to grow the sample of NIR spectra available to the community and use these NIR spectra to expand our physical understanding of transient explosions. Many of our spectra have been published in individual follow-up studies. This list included ASASSN-20hx/AT\,2020ohl \citep{Hinkle_2022}, SNe\,2020qxp/ASASSN-20jq \citep{Hoeflich_2021, Bose_2025},  
2020acat \citep{Medler_2023}, 2021csp \citep{Fraser_2021}, 2021yja \citep{Hosseinzadeh_2022}, 2022crv \citep{Dong_2024}, and 2020hvf and 2022pul \citep{OHora_2024}. 

Section~\ref{sec:Acqu} details the instruments used, their configurations, and the observing strategy. The data reduction methodology is presented in Section~\ref{sec:Redu}, followed by a summary of the sample statistics in Section~\ref{sec:Sample}. In Section~\ref{sec:spec_analysis}, we examine the global spectral features observed across different SN classifications. The highest-quality spectra for each classification are compared to template NIR spectra, and key spectral features identified in the best objects are analyzed in Sections~\ref{sec:template} and~\ref{sec:Analysis}, respectively. Finally, Section~\ref{sec:Summary} provides a summary of \textit{HISS} DR1.

\section{Data Acquisition} \label{sec:Acqu}
The observations presented in this work were obtained through a combination of $\sim156$~hours across 15 whole-nights and 3 half-nights on Keck-II using the Near-Infrared Echellette Spectrometer \citep[NIRES;][]{McLean_1998}, and $41.3$~hours over 9 nights on the NASA Infrared Telescope Facility (IRTF) using the SpeX medium-resolution spectrograph \citep{Rayner_2003}.
Keck-II/NIRES observations were obtained during semesters 2021A through 2023B, resulting in 76 spectra of 38 objects. In addition, 14 spectra of 10 objects were acquired using IRTF/SpeX during semesters 2021A and 2021B. IRTF data from semester 2022B onward will be included in future \textit{HISS} data releases, alongside observations from GNIRS on Gemini-North and additional Keck-II/NIRES data beginning in semester 2024A.

\subsection{Instruments and Observing Strategy}
Data was obtained using an ABBA dither pattern along the slit for both the NIRES and SpeX observations.
A $6$\arcsec offset between the A and B frames was used with NIRES and a $7.3$\arcsec offset for the SpeX observations. Bright objects were observed once with the dither pattern, while dimmer objects were observed multiple times to increase the signal-to-noise ratio. 
NIRES observations were all obtained using an individual A/B exposure time of 300\,s and multiple coadds depending on the magnitude of the target. The individual A/B exposure times of the SpeX observations was varied based on the apparent magnitude of the target to maximize the signal-to-noise. For SpeX, one coadd was used along with multiple cycles. SpeX targets were observed in either short wavelength cross-dispersion mode (SXD) or PRISM mode. 
In addition, spectra of standard A0V stars were obtained to correct the science exposures for atmospheric telluric effects and to perform flux calibration. Standard stars were observed near the target on the sky to ensure similar airmass conditions for accurate telluric correction.
The A0V standard stars were observed in an ABBA dither pattern, with a nod offset of $10$" for NIRES and $7.3$" for SpeX, using an exposure time of $3 - 10$ seconds per frame.

Finally, calibration data for both NIRES and SpeX were obtained each night to construct flat field images. An argon arc lamp was used to obtain images for wavelength calibration of the SpeX observations. Arc images were not obtained for the NIRES observations and wavelength calibration was done using NIR atmospheric lines. 

\subsection{Selection Criteria}
For a target to be selected for \textit{HISS} observations, it had to be spectroscopically classified as a SNe, and located within a distance of $D \leq 125$ Mpc. The only exceptions to this criterion were for exceptionally rare events such as the two interacting SNe, SN\,2021csp and SN\,2021kat, the SLSN\nobreakdash-I SN\,2021fpl, and the TDE ASASSN-20hx. These objects were luminous enough that their increased distance did not significantly affect the S/N of the observations. 
This limit on the distance was chosen to ensure that targets were sufficiently bright at late times, allowing for nebular phase follow-up observations to be obtained for many targets. 

\begin{figure*}
    \centering
    \includegraphics[width=\linewidth]{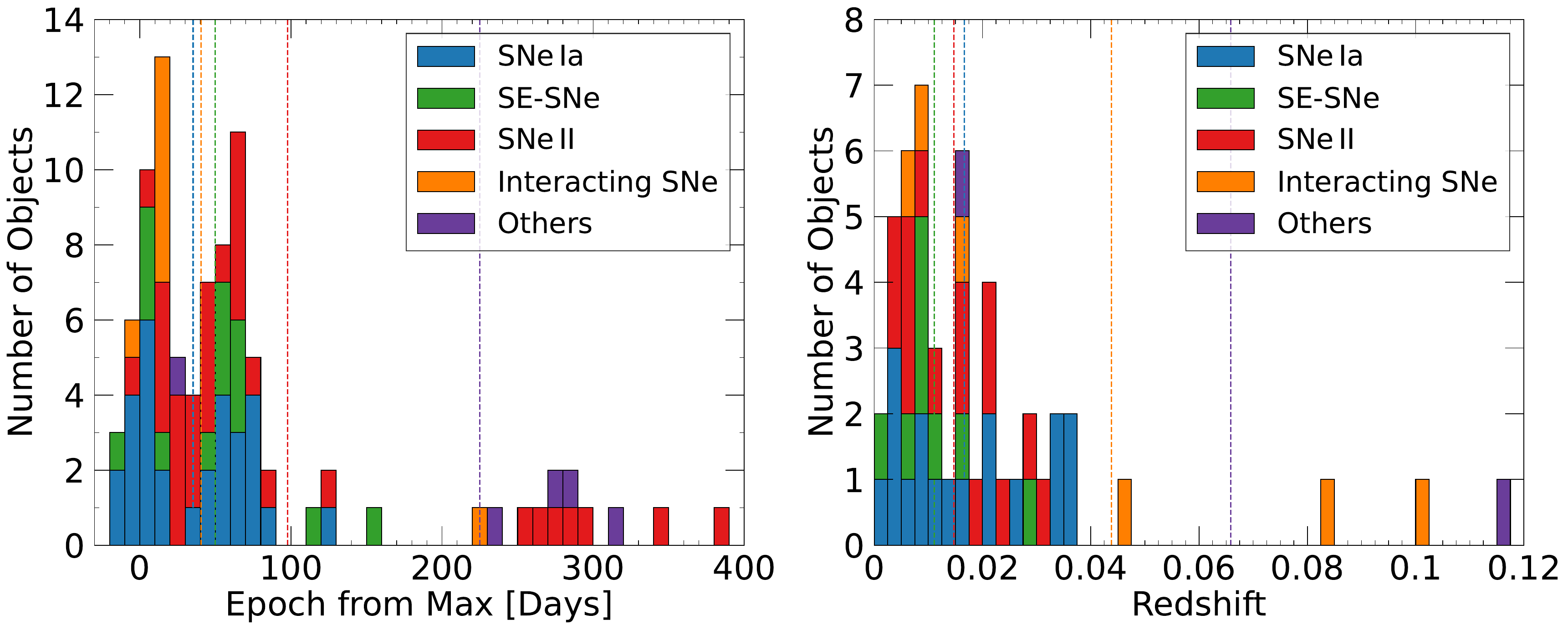}
    \caption{\textit{Left: }Phase distribution relative to epoch of maximum light in the $g/o$-bands for SNe\,Ia, SNe\,II, SE-SNe, Interacting SNe and other objects. \textit{Right:} Redshift distribution of the objects observed by \textit{HISS}. Vertical dashed lines denote the median phase and redshift for the different classifications observed in this data release.}
    \label{fig:phase+z_hist}
\end{figure*}

\section{Data Reduction} \label{sec:Redu}
\subsection{\textsc{Pypeit}}
Keck-II/NIRES spectra were primarily reduced using the open-source, \textsc{Python}-based spectroscopic reduction pipeline \textsc{Pypeit} \citep[][v1.14.0\footnote{https://pypeit.readthedocs.io/en/stable/index.html}]{Prochaska_2020a, Prochaska_2020b}, configured for Keck-II observations.

The raw images were first processed using the \textsc{pypeit\_setup} command, with the spectrograph set to NIRES, the configuration set to 'A', and background subtraction enabled. This command categorized the observations as either science, standard, or calibration. The science and standard images were also sorted into A/B pairs for background subtraction. \textsc{Pypeit} outputs these files into a data input file that was visually inspected to ensure that the A/B pairs were correctly assigned. Additional flags were added to the \textsc{Pypeit} input file to ensure that only one source was identified in the science and standard images. For dim sources, the \textsc{Pypeit} parameter \textit{snr\_thresh} was reduced to below 10, enabling the automatic extraction process to identify faint traces within the images.

Once all calibration, science, and standard files were correctly organized, the spectral extraction process was run through the \textsc{run$\_$pypeit} command. This command creates a master flat-field image using the input calibration files taken each night. The master flat fields were applied to both the science and standard images. The wavelength calibration function was constructed using the NIR OH sky lines that are visible in the science images. 
Flat-fielding was done using the dome flat calibration files, which were taken with the dome lamps turned off. Once the flat-field calibrations were applied to all science and standard files, the sources in the images were identified and extracted as \textsc{spec1d} files. During this process, each A/B pair of images was combined to construct a single image containing a positive and negative trace of the target. Each trace was extracted, and 1D/2D fits files containing the uncorrected spectrum of the object were constructed. This process was repeated for each A/B pair obtained during the night.

These files were flux-calibrated and corrected for telluric effects using several custom \textsc{Python} scripts that automated this process. To flux calibrate the science spectrum, a sensitivity function was constructed by dividing the A0V standard star spectrum, obtained close in time to the science exposure, by a model NIR spectrum of an A0V star adjusted for the atmospheric conditions at Maunakea. This sensitivity file was used to flux calibrate the science 1D spectra. All 1D spectra were subsequently coadded, and the orders were merged into a continuous spectrum. Finally, a telluric correction was applied to the science images using a telluric model created for Maunakea, completing the \textsc{Pypeit} reduction of the NIRES observations.

\subsection{\textsc{Spextool}}
Alongside the \textsc{pypeit} reductions, several spectra observations were reduced using the IDL-based software \textsc{Spextool}\footnote{\url{https://irtfweb.ifa.hawaii.edu/~spex/observer/}} \citep{Cushing_2004} and \textsc{xtellcor} \citep{Vacca_2003}. The first set of observations reduced using IDL are the low-luminosity/signal-to-noise NIRES targets. For these observations, the automated \textsc{pypeit} process struggled to determine the location of the trace in either one or all orders of the NIRES images during the extraction phase. The second group was the IRTF observations, as the \textsc{pypeit} code is not currently set up to reduce IRTF observations.

\textsc{Spextool} was run in both NIRES and SpeX modes as appropriate. The reduction process did not vary drastically between the two methods, the main difference being that IRTF/SpeX reductions use calibration arcs for the wavelength calibration. In contrast, the Keck-II/NIRES uses sky OH lines in the science images for the wavelength calibration. Wavelength calibration and flat field frames were constructed using the calibration images obtained during the night. Once the calibration files were constructed, the 1D spectrum was extracted by combining A/B pairs to obtain a positive and negative trace. The spectral profiles were constructed, and an aperture was chosen to capture most of the SN flux while limiting the background noise.  
The extracted A/B pairs were combined into a single extraction for the target and standard star using \textsc{xcombspec}. The combined standard star extraction was used to apply a telluric correction to the target extraction using \textsc{xtellcor} before the different orders were combined using the \textsc{xmergeorders} command. Any images with a low S/N or heavily contaminated by noise spikes were removed from the combining process to increase the total S/N of the output spectrum.

\section{Sample Statistics} \label{sec:Sample}
Our sample consists of 48 objects, including 47 supernovae and 1 TDE, with a total of 90 spectra covering multiple SN classifications. The details of the objects presented in this paper, such as their classification, date of maximum brightness, number of spectra, and redshift, are given in Table\,\ref{tab:SN_info}. The SN type for each object was obtained from the classification reports posted on the Transient Name Server (TNS)\footnote{\url{https://www.wis-tns.org/}}. 
Figure\,\ref{fig:Pie_charts} shows the total distribution of objects (left) and spectra (right) of this data release. For ease of analysis, the SNe have been grouped into one of five different types based on their spectral classification: SNe\,Ia, SNe\,II, SE-SNe, Interacting SNe, and other transients. In this data set, the largest number of individual objects observed were SN\,Ia, while the greatest proportion of spectra were obtained for SNe\,II.

\subsection{SNe Ia}
SNe\,Ia are the largest portion of individual objects, comprising $35.4\%$ of all objects presented in DR1, with 30 spectra of 17 SNe\,Ia, constituting $33.3\%$ of the total spectroscopic sample. Most of the SNe\,Ia are spectroscopically normal with two 91-bg-like SNe\,Ia (SNe\,2021qvv and 2022ihz), and one high-velocity SN\,Ia, SN\,2022hrs. 
All SNe\,Ia spectra obtained by \textit{HISS} in DR1 are presented in Figure\,\ref{fig:Ia_phot}.

\subsection{SNe II}
The second most common type of object observed, at $31.3\%$ of the total objects, was the hydrogen-rich core-collapse SNe\,II. For this data set, both Type IIb SNe (SNe\,IIb) and interacting Type IIn (SNe\,IIn) were excluded from the Type II group.  
SNe\,II comprise $36.7\%$ of the total spectra, with 33 spectra of 15 unique SNe\,II. The photospheric phase spectra, $t < +150$~days, are presented in Figure\,\ref{fig:II_phot} while the nebular phase spectra, $t > +150$~days, are shown in Figure\,\ref{fig:II_neb}. A spectrum of SN\,2023ixf was also obtained during semester 2023A. However, we do not add it to the sample here, as it will be presented in another study alongside optical and \textit{JWST} spectra in \citet{DerKacy_2025}.

\subsection{SE-SNe}
The SE-SNe sample contains $16.7\%$ of the total spectra and $15.6\%$ of the spectroscopic dataset with 14 spectra of 8 SE-SNe. This encompasses the hydrogen/helium-rich Type IIb, the hydrogen-poor/helium-rich Ib (SNe\,Ib), and hydrogen/helium-poor Ic (SNe\,Ic). The interacting Type Icn SN\,2021csp was excluded from the stripped envelope group. This dataset includes three SNe\,IIb, one SNe\,Ib and four SNe\,Ic. All SE-SNe spectra are presented in Figure\,\ref{fig:SE-SNe}.

\subsection{Interacting SNe}
The formation of narrow lines in the spectra of interacting SNe requires the SN ejecta run into a dense shell of CSM, which is independent of the exact explosion mechanism of the SN. Thus, objects in this class may range from SN\,Ia to SE-SN. DR1 contains 6 interacting SNe, $12.5\%$ of the total sample, comprised of one thermonuclear Type Ia-CSM SN (SN\,Ia-CSM), four hydrogen-rich Type IIn SNe (SNe\,IIn) including the peculiar SN\,2021aess, and the stripped envelope Type Icn (SN\,Icn) SN\,2021csp \citep{Fraser_2021}. In total, the interacting SN group has 8 spectra, $8.9\%$ of the total observations. The spectra of the SNe\,Ia-CSM, SNe\,IIn and SN\,Icn obtained by \textit{HISS} are presented in Figure\,\ref{fig:Interacting}.

\subsection{Other Objects}
Finally, there were two objects observed by \textit{HISS} that did not fit into any of the previous classifications. First, there is the TDE candidate ASASSN-20hx/AT\,2020ohl \citep{Hinkle_2022}. The four NIR spectra of ASASSN-20hx/AT\,2020ohl are shown in Figure\,\ref{fig:TDE_spec}. The other unique object observed by \textit{HISS} is SN\,2021fpl a superluminous Type I SN (SLSN-I), which is shown in Figure\,\ref{fig:SLSN_spec}. The spectra of these two objects make up $4.2\%$ of the total objects presented in this work and $5.6\%$ of all the spectra.

\subsection{Determining Maximum light}
The left panel of Figure\,\ref{fig:phase+z_hist} shows the distribution of spectral epochs relative to the time of $g$-band maximum light, $t_{max}$. To determine $t_{max}$ for the objects in this sample, the publicly available Zwicky Transient Facility (ZTF) photometry obtained between $-30$ and $+60$ days after detection was fit using a series of 6th- to 9th-order polynomials, with the best-fit orders used to determine $t_{max}$. Uncertainties in the date of maximum light were determined based on the cadence of the observations around peak light and the uncertainties in the fitting procedure. For those SNe that lacked comprehensive ZTF photometric coverage, additional sources, such as the ATLAS $c$- and $o$-bands or published optical data, were used to determine the epoch of maximum. Details on the source of additional photometry is given in Table\,\ref{tab:SN_info} Of the targets observed in our sample only the TDE AT\,2020ohl lacked both ZTF and ATLAS photometry. For this object, the date of maximum light was determined using ASAS-SN $g$-band data presented in \cite{Hinkle_2022}.

\subsection{Redshift Distributions}
The \textit{HISS} DR~1 redshift distribution is shown in the right-hand panel of Figure\,\ref{fig:phase+z_hist}. The mean redshift for SNe\,Ia, SNe\,II, SE-SNe, and interacting SNe are $z_\mathrm{mean} = 0.0168,\, 0.0138,\, 0.0111,\, \mathrm{and}\, 0.0408$ respectively. In addition, the TDE AT\,2020ohl and SLSN-I SN\,2021fpl possessed redshifts of $z = 0.01671$ and $z = 0.115$ respectively, with SN\,2021fpl possessing the highest redshift in this data set.

\subsection{Number of Observations per Object}
Most SNe were observed only once during the nights available between 2021 and 2024, but four SNe\,Ia, eight SNe\,II, three SE-SNe, and one interacting SN were observed more than once. Six SNe (SNe\,2022jzx, 2022mbg, 2022mit, 2022mww, 2022ojo, and 2022pul) were observed four times each over a four-day window towards the end of semester 2022B. Additionally, six other SNe were observed three times: SNe\,2020aatb, 2021tiq, 2021acnt, 2022jli, 2022crv, and 2022prr. Finally, four SNe were observed twice: SNe\,2020adow, 2021qvr, 2021agef, and 2022acko. 

\section{Identification and evolution of spectral features} \label{sec:spec_analysis}

Determining the presence and strength of spectral lines within NIR spectra is vital to understanding the properties of the evolving ejecta and the SN itself. To do this, we compare the smoothed spectra of SN with previous samples to identify common lines seen in the different classifications. All spectra were smoothed with a Gaussian filter using a sigma value of 3 and are shown in rest frame. 
Spectra have all been corrected for the Milky Way extinction assuming an $Rv = 3.1$ \citep{Schlafly_2011}, but they have not been corrected for host-galaxy extinction as exact extinction values are not known for the all objects. This does not significantly change our conclusions as the effect of extinction at NIR wavelengths is minimal \citep{Cardelli_1989}.

\subsection{SNe Ia}

Previous work on the NIR spectra of SNe\,Ia has shown a strong spectral homogeneity after maximum light \citep{Stanishev_2018, Muller-Bravo_2022, Lu_2023}. This homogeneity yields a significant agreement between the NIR spectral features of SNe\,Ia at similar epochs, with slight variations emerging depending on the degree of nuclear burning and \Nifs\ mass synthesized by the explosion \citep{Wheeler_1998}.
Throughout the evolution of SNe\,Ia, the NIR spectra are dominated by several lines associated with iron-group elements, primarily those of iron, nickel, and cobalt. The study of these lines enables the determination of the progenitor WD's structure and provides key insights into the explosion mechanism \citep{Hoflich_2002, Blondin_2023, Hoeflich_2021}.

\begin{figure*}
    \centering
    \includegraphics[width=0.75\linewidth]{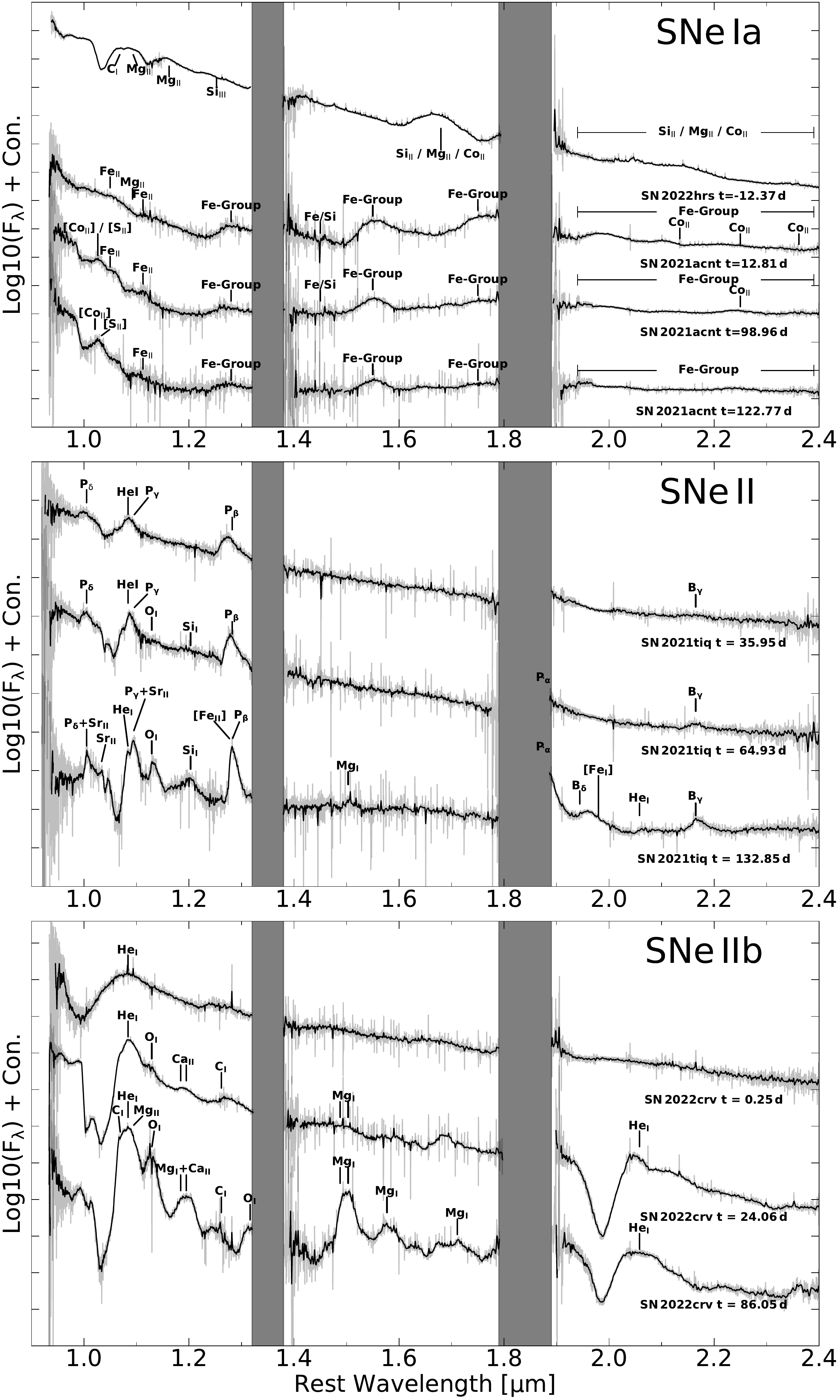}
    \caption{NIR spectra of four SNe representing a typical evolution of SNe\,Ia, SNe\,II and SNe\,IIb. These objects possessed the largest temporal range in DR1. Strong features are labeled based on previous studies of NIR SNe \citep{Marion_2009, Davis_2019, Shahbandeh_2022, Ashall_2019a}, with the label located at the rest wavelength of the line. }
    \label{fig:spec_temp}
\end{figure*}

The top panel of Figure\,\ref{fig:spec_temp} shows the NIR spectral evolution for the normal SNe\,Ia, SN\,2021acnt and SN\,2022hrs. The early time spectrum of SN\,2022hrs, obtained at $-12.37$~days, represents the earliest observation of a SN\,Ia in \textit{HISS} DR-1 and is used as the example for pre-maximum SNe\,Ia.
For the post-maximum SNe\,Ia, SN\,2021acnt possessed the largest temporal range in phase from the SNe\,Ia sample, ranging from $+12.81$~days to $+122.77$~days.

The spectra of SNe\,Ia, obtained before $t_{max}$, display a strong blue continuum with several P-Cygni features associated with intermediate-mass elements such as \MgII~$ 1.0927$~\micron, \MgII~$ 1.6787$~\micron, the weak \MgII~$ 2.1369$~\micron, \SiII~$ 1.6930$~\micron, \SiIII~$ 0.9800$~\micron, \SiII~$ 1.1737$~\micron, and \CoII~$ 1.6361$~\micron. Finally, several features seen redward of $2.0$~\micron\ are associated with Si, Mg, and Co lines.

After maximum light, the photosphere continues to recede and more of the \Nifs\ region becomes visible. This results in the emergence of several iron lines, e.g., \FeII~$ 1.0863$~\micron, \FeIII~$ 1.2786$~\micron, \FeIII~$ 1.2955$~\micron, \FeII~$ 1.5335$~\micron, \FeII~$ 1.6436$~\micron, and \FeIII~$ 1.6722$~\micron~\citep{Hsiao_2019}. Alongside the iron lines, several strong features associated with the cobalt \CoII~$ 1.5759$~\micron, $ 1.7772$~\micron, $ 1.7462$~\micron, $ 2.2205$~\micron, and $ 2.3613$~\micron\ lines are observed. The majority of these lines blend into broad emission features, especially those that lie between $1.3 - 1.8$~\micron\, (e.g. the $H$-band break, \citealp{Wheeler_1998}). These lines originate from allowed transitions of iron-group elements \citep{Wheeler_1998, Ashall_2019a, Ashall_2019b}. As SNe\,Ia evolve into the nebular phase, the allowed transitions fade from the spectra and forbidden lines, mainly associated with iron and cobalt, dominate the NIR spectrum \citep{Maguire_2018, Kumar_2023, OHora_2024}. 

\subsection{SNe II}

SNe\,II exhibit strong hydrogen features throughout their evolution. While the hydrogen features in SNe\,II are typically observed through the optical Balmer series lines ($n = 2$), strong features of hydrogen are also present in the NIR through the Paschen ($n = 3$) and Brackett ($n = 4$) series transitions. The Pa $\alpha$, $\beta$, $\gamma$, and $\delta$ lines are located $1.875$, $1.282$, $1.094$, and $1.005$~\micron, while the Br $\gamma$, $\delta$ and $\varepsilon$ lines are located at $2.166$, $1.944$, and $1.817$~\micron. The Pa$\alpha$, Br$\delta$, and Br$\varepsilon$ are all located in the telluric absorption band for low redshift objects and are typically only visible in the brightest objects \citep{Li_2025}. 

The middle panel of Figure\,\ref{fig:spec_temp} shows the NIR spectral evolution of the representative SN\,II from our dataset, SN\,2021tiq. During the early evolution, including both the rise and plateau phases, the spectra of SNe\,II are dominated by the recombination of hydrogen within the outer envelope resulting in strong P-Cygni features from the Paschen series in the NIR regime. Alongside the features associated with the different hydrogen series, the continuum of early time NIR spectra is influenced by the blackbody temperature of the ejecta, which evolves rapidly as the expanding ejecta cools.

As SN\,II evolve, photons from deeper in the ejecta escape, allowing for spectral features from heavier elements, such as \OI~$ 1.1290$~\micron\ and \SiI~$ 1.203$~\micron, to emerge. Additionally, helium features at $ 1.0830$~\micron\ and $ 2.0581$~\micron\ become visible after maximum light. The \HeI~$ 1.0830$~\micron\ line is blended with several other strong lines, including $\mathrm{Pa\gamma}$, Sr $\mathrm{_{II}}~ 1.092$~\micron\ and \MgII~$1.0927$\micron, while the \HeI~$ 2.0581$~\micron\ line, though isolated from other features, is significantly weaker. This lower intensity makes its identification challenging in spectra with low signal-to-noise. 

Finally, as the SN transitions into the nebular phase, the features evolve from P-Cygni profiles into pure emission lines. Forbidden transitions, such as [\FeII] $ 1.279$~\micron\ and [\FeI] $ 1.980$~\micron, become prominent in the spectra. These nebular phase emissions, when the photosphere has receded into the core of the ejecta, provide a way of determining the total mass and structure of the ejecta in a way that is not possible at earlier epochs \citep{Mazzali_2011, Dessart_2020}.

\subsection{SE-SNe}

The composition of the progenitor's outer envelope heavily influences the speed, strength, and type of the emerging features observed in spectra of SE-SNe. A SN\,IIb is observed if little stripping occurs and a thin hydrogen shell remains. While a SN\,Ib will occur if the full hydrogen envelope is removed, leaving an exposed helium envelope. Finally if both the hydrogen and helium shells are stripped prior to explosion a SNe\,Ic is observed. 

Among the sample of SE-SNe presented in this work, SN\,2022crv a H/He-rich SN\,IIb has the best temporal coverage. The identified spectral lines of SN\,2022crv are shown in the bottom panel of Figure\,\ref{fig:spec_temp}. Another SE-SN included in DR1, SN\,2022jli (Type Ic), was observed multiple times. However this occurred in close succession and does not provide the temporal coverage of SN\,2022crv.
Although SN\,2022jli does not possess as large a temporal baseline as SN\,2022crv, it was a unique SN\,Ic that displayed periodic undulations in its light curve with a period of $\sim 12.5$~days \citep{Moore_2023, Cartier_2024} which arises as material is accreted onto the supernova remnant from a binary companion star \citep{Moore_2023, Chen_2024, King_2024}.

Hydrogen- and helium-rich SNe\,IIb such as SN\,2022crv exhibit strong Paschen lines, along with helium features at $1.0831$~\micron\ and $2.0580$~\micron. However, when the SN transitions to the nebular phase --- sometimes as early as  40 and 60 days after the explosion in the NIR --- the Paschen lines fade and helium lines dominate the spectrum. At this stage, additional spectral features emerge and the spectra of SNe\,IIb become more similar to SNe\,Ib. These features are associated with elements from the inner ejecta, including strong spectral lines such as the  \CI~$ 1.2614$~\micron, \OI~$ 1.1290$~\micron\ and $ 1.3165$~\micron, \CaII~$ 1.1839$~\micron\ and $ 1.1950$~\micron, as well as \MgII~$ 1.4878$~\micron\ and $ 1.5033$~\micron\ \citep{Bufano_2014, Shahbandeh_2022}.

NIR spectra provide the best insight into the helium shell through the \HeI~$1.0831$~\micron\ and $2.0580$~\micron\, lines which directly trace the structure of the helium envelope. These lines are stronger and more isolated, respectively, than any optical helium line \citep{Lucy_1991}. Analysis of these features in the late-time spectra ($t_\mathrm{exp} > +100$~days) for three SNe\,IIb has found a distinct difference that emerges in the helium features. In SN\,2008ax \citep{Pastorello_2008} and 2020acat \citep{Medler_2022}, helium features were found to develop flat-topped profiles at late times, whereas SN\,2011dh \citep{Arcavi_2011} exhibited only a sharp emission feature during the nebular phase \citep{Medler_2023}. A tentative connection has been suggested between the morphology of late-time helium emission and the compactness of the progenitor’s outer envelope \citep{Medler_2023}. A larger sample of late-time NIR spectra of SNe\,IIb is required to confirm the link between progenitor structure and evolution of the helium shell.

In hydrogen-poor, helium-rich SNe\,Ib, the most prominent features in the NIR spectrum are the \HeI~$1.0830$~\micron\ and $2.0581$~\micron\ features \citep{Shahbandeh_2022}. 
As these SNe evolve, additional carbon lines emerge, including \CI~$ 1.0693$~\micron\ and $ 2.1259$~\micron, both of which blend with helium features \citep{Shahbandeh_2022}. In addition, SNe\,Ib display features similar to SNe\,IIb with lines of oxygen, magnesium, and calcium present in their NIR spectra. The magnesium lines at $\sim 1.5$~\micron\ often appear as a single blended feature; however, in high-resolution spectroscopy, they can be resolved into individual peaks superimposed on a broad emission profile \citep{Shahbandeh_2022}. At later times, several lines associated with iron and cobalt emerge. These include \FeIII~$ 1.6672$~\micron, \FeIII~$ 1.6722$~\micron, \CoII~$ 1.6361$~\micron, and \CoII~$ 1.7239$~\micron. These features are typically highly blended with other lines, making their identification challenging.

Finally, the most stripped type of SE-SNe in the sample are the hydrogen/helium-poor SN\,Ic. The removal of the hydrogen and helium envelopes prior to explosion should prohibit the spectra of SNe\,Ic from displaying H/He lines. However, stellar evolutionary models have trouble fully removing the helium envelope of SNe\,Ic progenitors, as such it is possible that SNe\,Ic's possess a thin envelope of helium present at the time of explosion \citep{Hachinger_2012, Teffs_2020, Williamson_2021}. Detecting helium in SNe\,Ic at optical wavelengths is difficult, due to the high ionization potential of the lines \citep{Lucy_1991} and strong blending effects. However, the two NIR \HeI\ lines can be used as diagnostic tools to determine the thickness of the helium envelope in SNe\,Ic \citep{Hachinger_2012}.

Typically SNe\,Ic display a strong feature located in the region of the \HeI~$ 1.0830$~\micron\ that is a blend of the \CI~$ 1.0693$~\micron\ and \MgII~$ 1.0927$~\micron\ lines, with potential contamination from the \SI~$ 1.0457$~\micron\ line \citep{Taubenberger_2006}. At late times, these SNe exhibit similar emission features to SNe\,IIb and Ib, displaying emission features associated with oxygen, calcium, magnesium, iron, and cobalt lines \citep{Hunter_2009, Mazzali_2010}.

\subsection{CSM Interaction}

\begin{figure*}
    \centering
    \includegraphics[width=\linewidth, trim={1cm 0.5cm 3cm 1cm},clip=True]{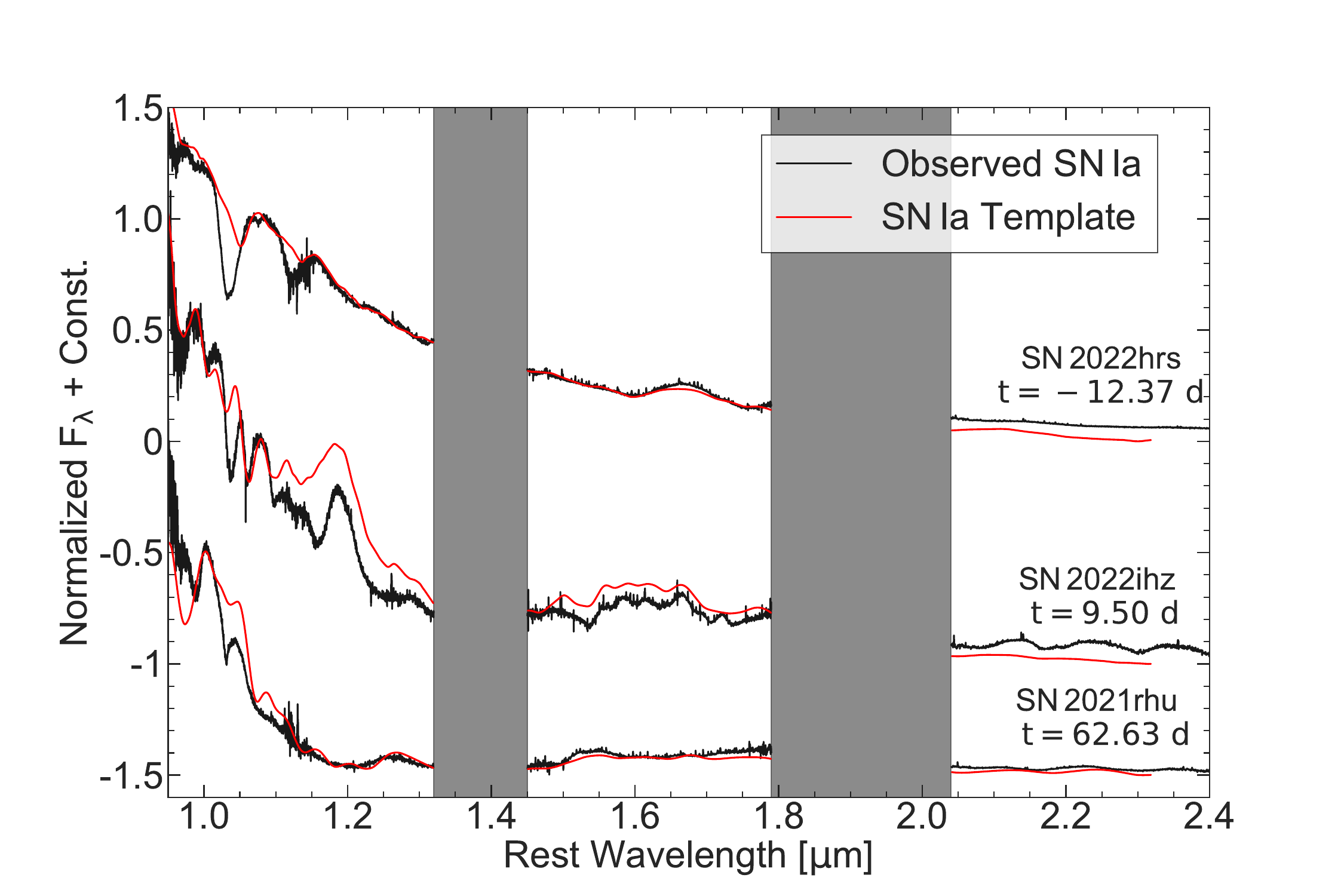}
    \caption{Comparison of SN\,2021rhu, SN\,2022hrs, and SN\,2022ihz to the spectral template constructed using the data presented in \citep{Lu_2023}. Spectra are shown in rest wavelength, and the gray regions denote telluric regions. The $s_{BV}$ values used in constructing the template spectra were obtained by matching the template to SNe spectra at each epoch. As such, the strength features do not fully match between the template and observed spectra. However, the locations of most spectral features are consistent, and detailed spectral modeling is required to accurately reproduce the strength of spectral features.}
    \label{fig:SNIa-temp}
\end{figure*}

The NIR spectra of SNe\,Ia-CSM and SNe\,IIn are typically populated by strong, narrow hydrogen lines from the Paschen and Brackett series, as shown in Figure\,\ref{fig:Interacting} \citep{Schegel_1990, Filippenko_1997, Silverman_2013}. The narrow emission lines from the Pa$\beta\,1.282$~\micron, and Pa$\gamma\,1.094$~\micron\ lines are clearly seen in SNe\,2020aekp at $+211.74$~days, SN\,2021aess at $+28.86$~days and SN\,2022prr at $\sim+ 24.68$~days, in addition the Pa$\alpha\,1.875$~\micron\ line is observed in SN\,2020aekp. The average full width half maximum (FWHM) of the Pa$\beta$ line for these three SNe is $\mathrm{FWHM} = 119 \pm 20$~\kms. The velocities determined from fitting to the narrow emission lines, along with the properties of the SN, can be used to derive the mass-loss history of the progenitor \citep{Kotak_2004, Inserra_2016, Yang_2023}. For example, the strong Pa$\beta$ emission feature in SN\,Ia-CSM 2020aekp exhibits a FWHM of $102 \pm 15$~\kms. Using equations $3 - 6$ from \citet{Yang_2023} a mass-loss rate $\dot M \approx 10^{-3} \, \mathrm{M_{\odot}\,yr^{-1}}$ was found for SN\,2020aekp. The mass-loss rate was calculated assuming a typical expansion velocity for SNe\,Ia of $\sim 7000$~\kms, a spherical distribution of the CSM and expanding ejecta, and a conversion efficiency between kinetic energy and radiation of $\sim 20\%$ \citep{Yang_2023}. This mass-loss rate results in a CSM mass for SN\,2020aekp of $M_{CSM} \approx 1$~\Msun. 
In contrast, the interacting SE-SNe, such as SN\,2021csp, display strong lines of helium, oxygen, carbon, or nitrogen \citep{Fraser_2021, Gal_Yam_2022, Perley_2022, Schulze_2024}. 
As the density of the CSM changes, the strength of the resulting narrow emission lines varies. Some interacting SNe have been found to display several epochs of interaction resulting from multiple shells of CSM formed by eruptive mass loss episodes from the progenitor before explosion \citep{Nyholm_2017, Hiramatsu_2024}.

The interaction between the SN ejecta and CSM that powers interacting SNe is not a continuous process and strongly depends on the geometry of the CSM \citep{Smith_2017, Kurfurst_2019}. The spectra of interacting SNe may go through period where no narrow emission features are observed within the NIR spectra, depending on the epoch of observation and the location of the forward shock relative to the dense CSM. This is seen in SNe\,2021kat, 2021foa, and 2021csp, all of which lack strong narrow emission features. The lack of there narrow emission features at the observed epochs can be used to constrain the radius of the CSM and thus the mass loss rate of the progenitor \citep{Smith_2017}.

\begin{figure*}
    \centering
    \includegraphics[width=\linewidth, trim={1cm 0.5cm 3cm 1cm},clip=True]{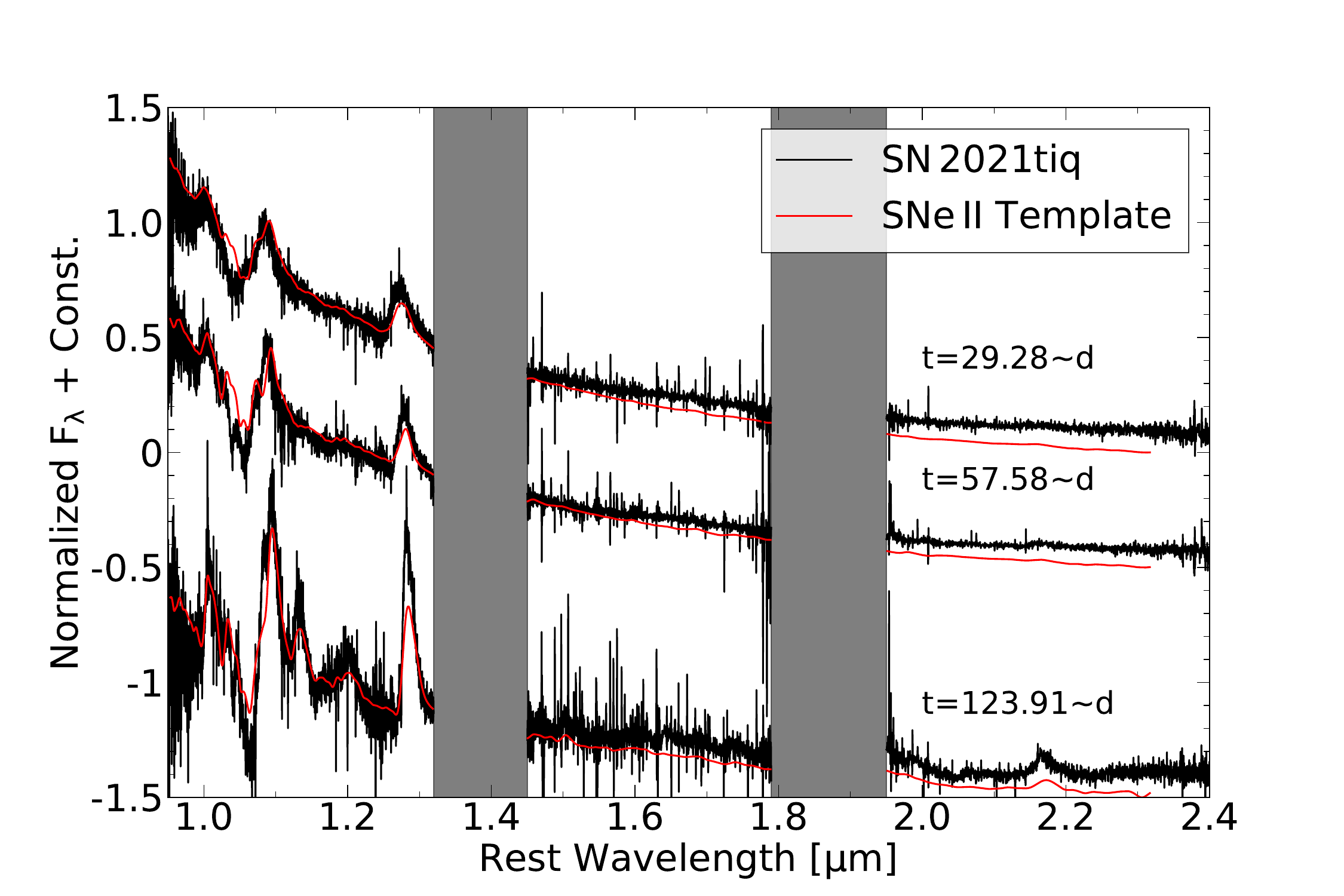}
    \caption{Similar to Figure\,\ref{fig:SNIa-temp} but for SNe\,II. Comparison of SN\,2021tiq, at three distinct epochs after explosion, with spectral template of SN\,II obtained from \citet{Davis_2019}. All spectral features observed in the template spectra can be found within the spectra of SN\,2021tiq.}
    \label{fig:SN_II_temp}
\end{figure*}

\subsection{ASASSN-20hx/AT 2020ohl}
The NIR spectra of nuclear transients like TDEs are currently poorly understood, likely due to the dilution of transient features by the much stronger host-galaxy flux in the NIR. However, there are several NIR features that TDEs are expected to exhibit. The spectra of TDEs are commonly dominated by a hot (T $\sim 20,000 \mbox{ -- } 50,000$ K) blackbody, but some reprocessing models predict a spectral flattening at NIR wavelengths \citep{Roth_2016, Lu_2020}. Additionally, several TDEs have shown strong high-ionization coronal lines in their optical \citep[e.g.,][]{Wang_2012, Clark_2024} and NIR \citep{Onori_2022} spectra. These lines provide unique probes of the extreme UV and soft X-ray emission from these events. These coronal lines can therefore be used to understand the full radiated energy from these accretion flares and can also be compared to the well-studied behavior of NIR coronal lines in active galactic nuclei \citep{Lamperti_2017}.

Although the sample of TDEs with NIR spectra is small, the existing data shows interesting diversity. Some TDE NIR spectra are featureless \citep[e.g., ASASSN-23bd][]{Hoogendam_2024_TDE}, yet other TDEs, like ASASSN-22ci \citep{Hinkle_2024} and AT2017gge \citep{Onori_2022}, show broad \HeI~$1.083$~\micron\ emission. Some TDEs, like AT2017gge, additionally show broad emission from the Paschen series and narrow coronal line emission \citep{Onori_2022}. 

As ASASSN-20hx/AT 2020ohl is the only object in this sample that is neither the thermonuclear destruction of a WD nor the collapse of a massive star, it displays drastically different features from those previously discussed. The NIR spectra of ASASSN-20hx are blue and featureless, lacking any of the broad Balmer and/or He features seen in TDE optical spectra or the narrow coronal line emission seen in some TDEs and common in AGNs. The first two spectra of AT\,2020ohl, obtained at $+225.79$ and $+262.64$~days are presented in \citet{Hinkle_2022}. The additional late-time spectra continue this trend. This hot blackbody-like emission, persisting several hundred days after peak light, is a hallmark of accretion-powered transients.

\begin{figure*}
    \centering
    \includegraphics[width=\linewidth, trim={1.4cm 0.5cm 3cm 1cm},clip=True]{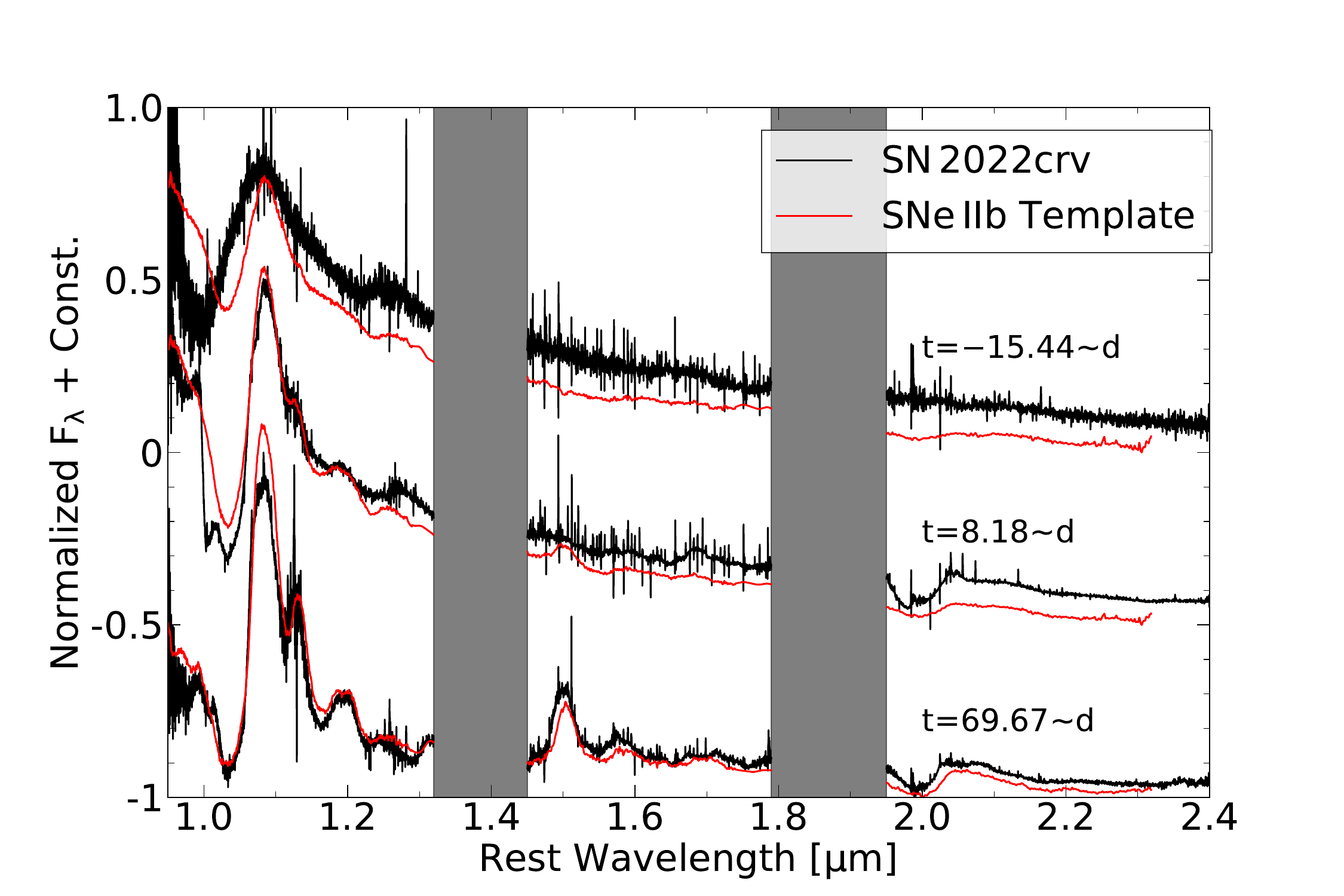}
\caption{Similar to Figure\,\ref{fig:SNIa-temp} but for SNe\,IIb. Comparison of SN\,2022crv with the spectral template of He-rich SE-SNe presented in \citet{Shahbandeh_2022}. The template spectra reproduce the spectral features present in the spectra of SN\,2022crv. The initial spectra of SN\,2022crv, taken at $-15.24$~days, are compared to the earliest template at $-5$~days, resulting in a difference in ejecta velocity.}
    \label{fig:SE-SN_temp}
\end{figure*}

\subsection{SLSN-I 2021fpl}

Due to the high redshifts of SLSNe-I, they are rarely observed in the rest-frame NIR. Only a handful of SLSNe-I have been observed at NIR wavelengths, including SN\,2012il \citep{Inserra_2013}, LSQ14an \citep{Jerkstrand_2017}, SN\,2015bn \citep{Nicholl_2016}, and Gaia16apd \citep{Yan_2017}. SN\,2021fpl, which we observed at $+21.61$~days, displayed two strong features, located around $\sim 0.94$~\micron\ and $\sim 1.1$~\micron. The $\sim 1.1$~\micron\ feature has been associated with the \HeI~$1.083$~\micron\ respectively and has been identified in multiple SLSNe-I \citep{Inserra_2013, Nicholl_2015, Yan_2017} and other hydrogen-poor CC-SNe, as discussed above. The identification of the line responsible for the $0.94$~\micron\ feature is more ambiguous.\citet{Yan_2017} associated the feature in GAIA16apd with the \CII~$0.9234$~\micron\ line. However, the feature in SN\,2012il and SN\,2015bn was identified as a \MgII\ line \citep{Inserra_2013, Nicholl_2016}, potentially the \MgII~$0.9218$~\micron\ or $0.9393$~\micron. \citet{Inserra_2013} also identified the \MgII~$1.5024$~\micron\ line within the NIR spectra of SN\,2012il. The lack of a clear features around $1.5$~\micron\ in SN\,2021pfl suggests the $0.92$~\micron\ feature is likely associated with the \CII\ line and not \MgII. 

\section{Template comparison} \label{sec:template}

Comparing the observed spectra to SNe templates enables differences in line strength, profile, and shape to be identified, providing information about the progenitor's properties and allowing unique objects to be easily identified. For example, the line strength of hydrogen/helium can determine the extent of stripping in SE-SNe \citealt{Gilkis_2022, Ergon_2022}. Therefore, we compare our best observed SNe to pre-existing spectral templates.   

The closest phase template spectra for each type of SN were matched to the epoch relative to $t_{max}$ of the observed spectra. 
The continua of the SNe and template spectra were fit using the \textsc{fit\_continuum} package of the \textsc{python} software \textsc{specutils}.
The template spectrum was corrected to match the continuum of the SN spectrum. This allows the line profiles of the templates to be directly matched to those of the sample SN, enabling any differences in spectral features to be identified. All spectra, object and template, have been normalized at $1$~\micron.

The templates of the SNe\,Ia and CC-SNe match the observed \textit{HISS} spectra. As expected, several differences are observed between the templates and the actual spectra, which are discussed below. Despite this, the agreement between the templates and our spectra demonstrates both the quality of the data reduction process and the power of using high-quality templates to discern the unique properties of new SNe.

\subsection{SNe Ia Template Comparison}

Comparing the early time spectra of SNe\,Ia, especially those obtained within hours to days of the explosion, to spectral templates provides a unique look into their diversity \citep{Hoogendam_2025a}.
Figure\,\ref{fig:SNIa-temp} shows a comparison of three normal SNe\,Ia at different phases relative to the templates from \citet{Lu_2023}.
Constructing template spectra of SNe\,Ia, requires knowledge of both the phase relative to peak light and the shape of the SN's light curve \citep{Lu_2023}. The shape of SNe\,Ia light curves can be characterized by the $s_{BV}$ value: the time of maximum $(B-V)$ color divided by $30$~days \citep{Burns_2014}. For the SNe shown in Figure\,\ref{fig:SNIa-temp}, it was not possible to determine the $s_{BV}$ value as they lack $B$- and $V$-band coverage. Instead, the $s_{BV}$ value was varied between 0 and 1.5 to construct an array of template spectra for each epoch. The template spectra that most accurately reproduce the features seen in the observed spectra are shown in Figure\,\ref{fig:SNIa-temp}. 

While the templates were able to reproduce the majority of the NIR features, some spectral features do not fully match in terms of line strength or the velocity of the absorption features. However, as template comparisons are used to determine the major differences between observations and the typical SN\,Ia, this is not an issue. This discrepancy is evident in the earliest spectrum shown in Figure\,\ref{fig:SNIa-temp} for SN\,2022hrs. The absorption minima observed in the features of SN\,2022hrs are much faster than those of the template spectrum. There are two reasons for the difference in the line velocities of the spectral features. The first cause is that the SN\,2022hrs spectrum is $\sim 7$ days earlier than the first template. The second cause is SN\,2022hrs possessed significantly faster ejecta than more typical SNe\,Ia \citep{Benetti_2005, Risin_2023}. The late-time NIR spectra of SNe\,Ia presented in this work display a high degree of homogeneity and match well with the template spectra. However, at earlier times, especially within the hours and days after the explosion, there seems to be a larger diversity observed in SNe\,Ia \citep{Hoogendam_2025a}. An increased number of NIR spectra of SNe\,Ia during these early phases is critical to quantifying the homogeneity of the outermost regions of the ejecta and improving early time spectral templates.

\subsection{SN II Template Comparison}

Among the Type II supernovae in our sample, SN\,2021tiq has the most extensive temporal coverage, with observations obtained at $+29.28$, $+57.58$, and $+123.91$ days relative to $t_{max}$. This large temporal coverage makes it the ideal SN from our sample to compare to the spectral templates constructed by \citet{Davis_2019}. The comparison between the template spectra and SN\,2021tiq are shown in Figure\,\ref{fig:SN_II_temp}. The first two epochs were obtained during the plateau phase of SN\,2021tiq, where the hydrogen Paschen features dominate the spectra. The Paschen features match the template spectra well. As SN\,2021tiq evolved, at $+57.58$~days, additional lines emerge, predominantly around the $1$~\micron\ feature. In addition, the absorption profiles decline in width as the velocity of the emitting region decreases. Finally, in the last epoch, emission lines from heavier elements, such as iron and strontium, fully emerge within the spectrum. 

One of the major differences between the template and SN\,2021tiq is seen in the $1.3$~\micron\ emission feature observed at day $+123.91$. In the template, there is a single feature with a smooth emission peak originating from the Pa$\beta$ line. However, this feature is commonly observed as a blend of the Pa$\beta$ and [\FeII] $ 1.279$~\micron\ lines. In addition, the hydrogen features in the spectrum of SN\,2021tiq at day $+123.91$ display red shoulders in the peak of their profiles, which may indicate the presence of dust forming within the ejecta, interaction with the surrounding CSM \citep{Chugai_2007}, or an asymmetric effects within the ejecta \citep{Bose_2015}. This feature has been observed in the late time of several, but not all, SNe\,II \citep{Bose_2015}. We therefore encourage the transient community to target SNe\,II beyond $+100$~days in the NIR, which is under-sampled compared to nebular optical spectra. These late-phase observations are essential for dissecting forward-shock/CSM interactions and constructing a high-cadence time series following the formation of molecules and dust within the system.

\subsection{SE-SNe Template Comparison}

Comparisons between SN\,2022crv and the template for He-rich SE-SNe created by \citet{Shahbandeh_2022} highlight the major differences between the templates and new SNe\,IIb events. The three epochs of SN\,2022crv, taken at $-15.24, +8.38$ and $+69.87$ days after $t_{max}$ compared to the template spectra are shown in Figure\,\ref{fig:SE-SN_temp}. The initial observation of SN\,2022crv, at $t = - 15.44$~days, was $\sim 2.8$~days after explosion and is one of the earliest SE-SNe observed in the NIR \citep{Dong_2024}. This resulted in a phase discrepancy between SN\,2022crv and the template spectra, which was created at $-5$ days prior to $B$-band peak, by $\sim 10$~days. As such, the velocity profile of the \HeI~$ 1.0830$~\micron\ feature does not match the template; however, the other line profiles are consistent between the two spectra. As SN\,2022crv evolved, the absorption feature associated with the $Pa\gamma 1.094$~\micron\ and \HeI~$ 1.0830$~\micron\ lines splits into two distinct features. This split originates from the presence of a high velocity thin hydrogen shell, at $\sim 15000$~\kms\ at day $8.38$, overlapped with the helium feature, with a velocity of $\sim 9500$~\kms. This is in agreement with the velocity evolution observed in the optical lines of SN\,2022crv \citep{Gangopadhyay_2023, Dong_2024}. This high-velocity hydrogen feature declines in strength as hydrogen fades from the spectrum \citep{Dong_2024} such that by day $69.87$ the spectrum of SN\,2022crv aligns significantly with the He-rich SE-SNe template spectrum. Increasing the number of young NIR spectra of SE-SNe will significantly improve the quality of early-time template spectra.

\section{Unique NIR Analysis} \label{sec:Analysis}

NIR spectra contain a wealth of information about the physics of SNe explosions. Here, we highlight several examples of these interesting features including the shape of the $1$~\micron\ feature in SNe\,Ia, the structure of nebular emission lines in SNe\,II, the presence of the Carbon Monoxide first overtone in CC SNe, and the strength of the Helium $1$~\micron\ absorption feature in CC SNe.

\begin{figure}
    \centering
    \includegraphics[width=\linewidth, trim={1.15cm 0.5cm 1.35cm 0.5cm}, clip=True]{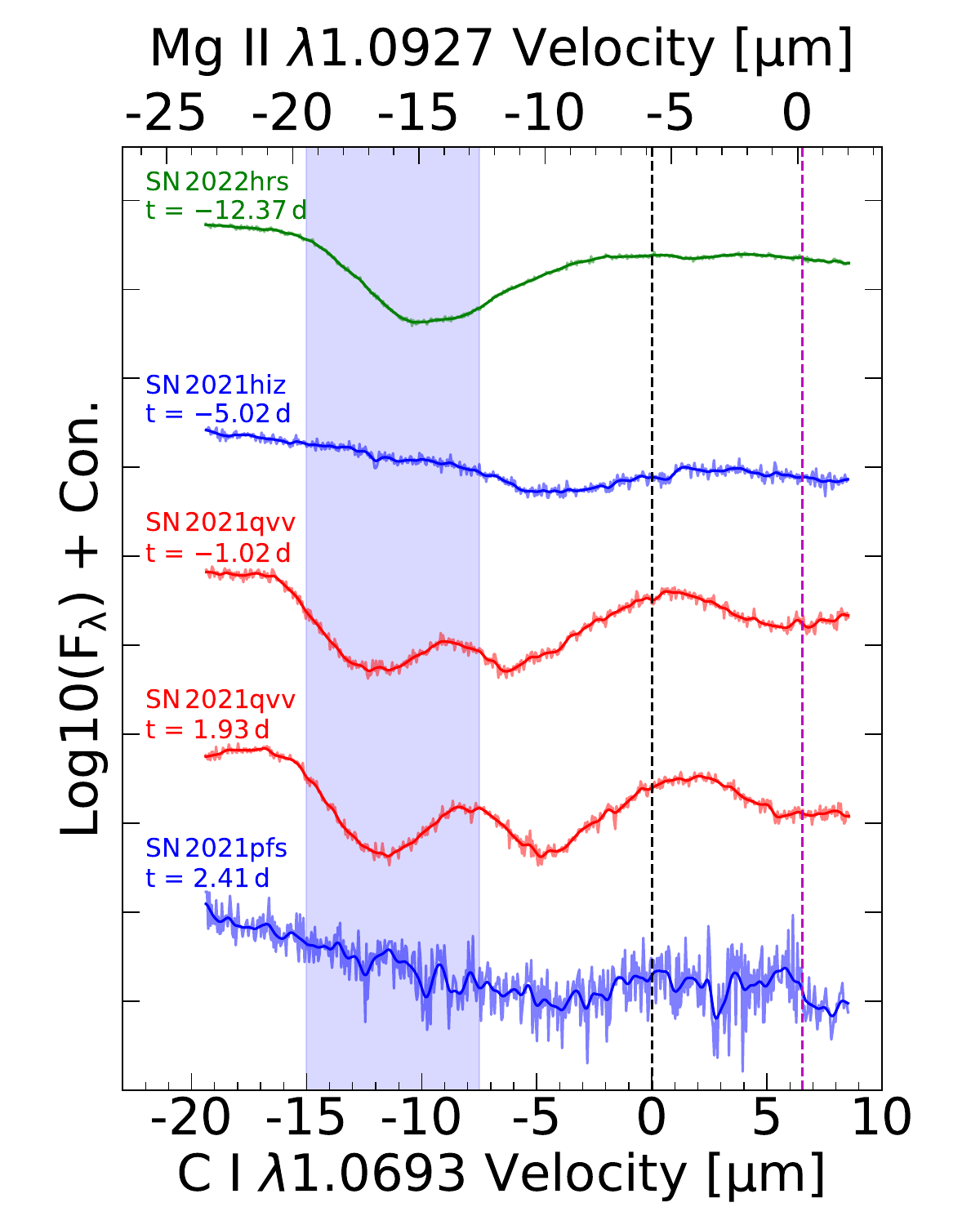}
    \caption{Structure of the \CI~$1.0693$~\micron\ line observed in the pre-peak NIR spectra of SNe\,Ia. The rest wavelength of the \CI~$1.0693$~\micron\ and \MgII~ $1.0927$~\micron\ lines are given by the dashed vertical lines. The spectra are divided into three groups and color-coded according to the structure of the absorption feature. The first group includes SNe that lack any absorption features (blue), the second group displays a single ``V" absorption feature in the $1$~\micron\ region (green) associated with high velocity \MgII~ $1.0927$~\micron\ and finally are spectra displaying a ``W" feature (red) associated with the presence of magnesium and unburnt carbon in the ejecta.}
    \label{fig:SN_Ia_Cw}
\end{figure}

\subsection{The C I 1.0693 micron feature in SNe Ia}
Carbon is the only fully unburnt element that is present in the spectra of SNe\,Ia. Detecting it and following its evolution has direct implications for the progenitor and explosion scenario \citep{Hoeflich_2002, Fink_2010, Pakmor_2012}. While \CII\ is visible in the optical, it is on top of the emission component of the strong \SiII\ $0.6355$~\micron\ P-Cygni profile \citep{Thomas_2011, Folatelli_2012} and requires a higher ionization energy than the \CI\ lines to be excited. However, the NIR \CI\ $ 1.0693$~\micron\ line is significantly stronger and more isolated than the \CII\ line, allowing for easier identification. To search for \CI~$ 1.0693$~\micron\ in our data, we examined the shape of the $1$~\micron\ feature within the spectra of all SNe\,Ia obtained within three days of $t_{max}$ or earlier. Four SNe\,Ia in the \textit{HISS} DR1 were observed prior to $+3$~days: SN\,2022hrs at $-12.37$~d, SN\,2021hiz at $-5.02$~d, SN\,2021qvv at $-1.02$ and $1.93$~d, and SN\,2021pfs at $2.41$~d.
The NIR spectra of these SNe are presented in Figure\,\ref{fig:SN_Ia_Cw}. The spectra are shown in velocity space relative to the strong \CI~$ 1.0693$~\micron\ and presented in time order. These spectra display one of three features within the $1.0$~\micron\ region. First, there are SNe that lack any high velocity absorption feature within the $1.0$~\micron\ region, known as ``flat'' SNe\,Ia. Secondly the SNe that possess a single strong ``V''-shaped absorption feature, attributed to the \MgII~$1.0927$~\micron\ line. Finally, the double absorption features SNe, displaying a ``W''-like shape in the spectra, are best explained by a combination of the \CI~$ 1.0693$ \citep[e.g.,][]{Hsiao_2015, Hoogendam_2025a} and \MgII~$1.0927$~\micron. 

We find a clear detection of \CI~$ 1.0693$~\micron\ in SN\,2021qvv, which is consistent with previous results that lower luminosity SNe\,Ia have less efficient burning and contain a large amount of unburnt material \citep{Hoeflich_2002, Hsiao_2015, Wyatt_2021, Pearson_2024, Hoogendam_2025a}. The broad ``V'' shape that is seen in SN\,2022hrs may indicate the presence of high velocity magnesium, which agrees with the ejecta velocity derived from optical lines \citep{Risin_2023}. Finally, two of the young SNe\,Ia in our sample are ``flat'', SNe\,2021hiz and 2021pfs, which lack either high velocity magnesium or carbon absorption features. SN\,2021hiz, at $t = -5.02$~days, and SN\,2021pfs, $t = +2.41$~days, show weak \MgII~$1.0927$~\micron\ absorptions, located at $\sim 11,000$~\kms roughly $2 - 4 \times 10^3$~\kms\ slower than either the ``V'' or ``W'' SNe\,Ia at similar epochs.
Understanding the diversity of $1.0$~\micron\ feature in pre-maximum SNe\,Ia will be important to differentiate between different explosion models \citep{Hoeflich_2002, Kasen_2009, Hsiao_2015}. A larger sample of pre-maximum NIR spectra of SNe\,Ia is required to explore this parameter space. 

\begin{figure*}
    \centering
    \includegraphics[width=\linewidth]{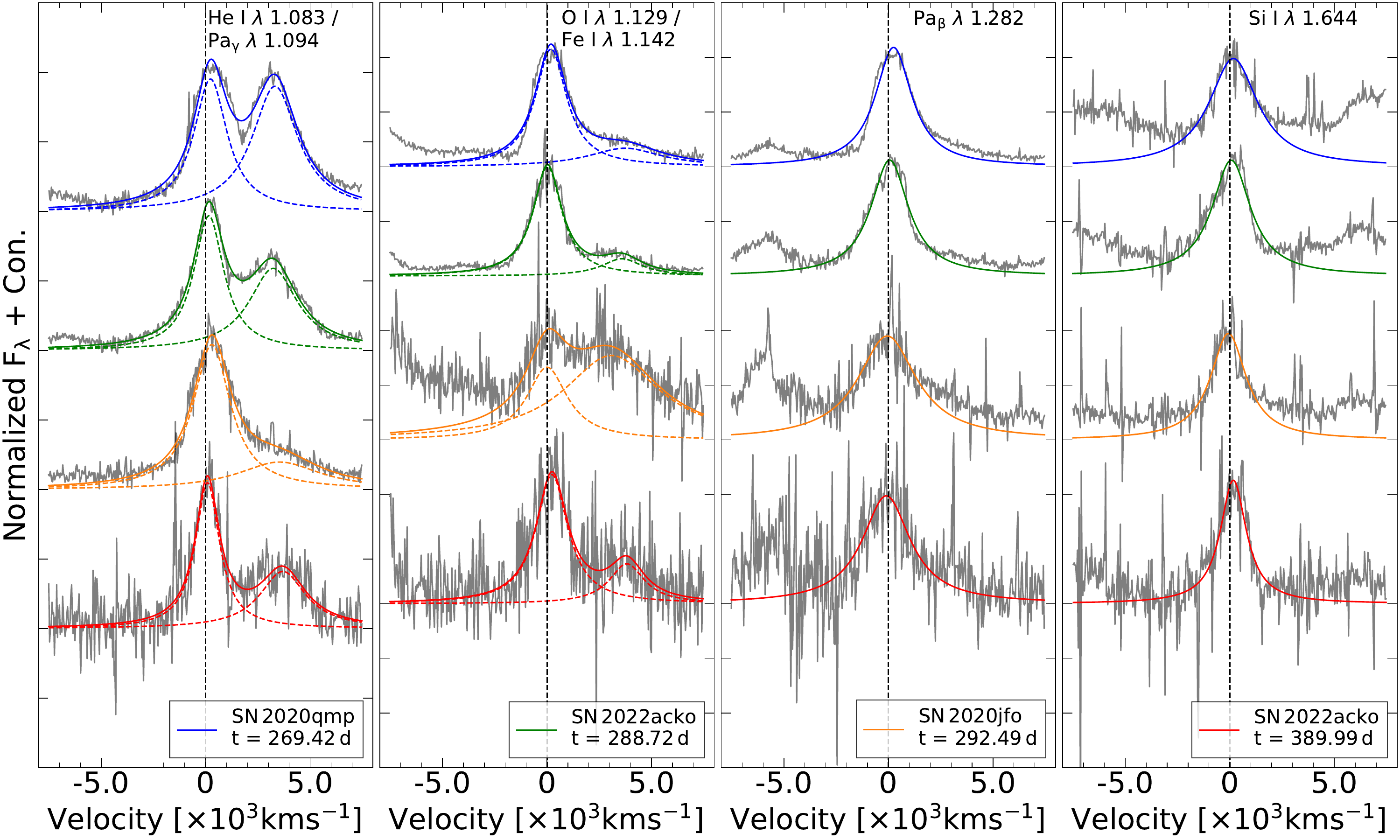}
    \caption{Nebular-phase emission features of the \HeI~$1.083$~\micron\ / Pa$\gamma\,1.094$~\micron, \OI~$1.129$~\micron\ / \FeI~$1.142$~\micron, Pa$\beta\,1.282$~\micron, and \SiI~$1.644$~\micron\ lines of the three SNe\,II SN\,2020jfo, SN\,2020qmp and SN\,2022acko. Features were fitted with either a single or double Lorentzian function to determine the FWHM of the emission lines.}
    \label{fig:SN_II_Neb_Line_Fits}
\end{figure*}

\subsection{Nebular SNe II Emission Lines}

As the photosphere of a CC-SN recedes through the ejecta, more of the inner material becomes visible as the SN transitions into the nebular phase ($t \geq 150$~d) \citep{Jerkstrand_2017}. During this phase, the spectrum of SNe is powered by the radioactive decay of \Cofs\ and is dominated by forbidden emission lines as the ejecta has become optically thin. 
Late-time spectroscopic observations of CC-SNe provide the best way of determining the structure and properties of the inner ejecta, which is directly related to the attributes of the progenitor. However, obtaining observations of SNe at this phase is difficult due to the intrinsically low luminosity during this period. As such, late-time nebular spectra are not typically obtained for SNe\,II, which is exacerbated in the NIR. 

Understanding the properties of the nebular phase in SNe\,II requires that we analyze the line profiles of the strongest lines within the NIR data presented in this sample. As such we look at the \HeI~$1.083$~\micron\ / Pa$\gamma\,1.094$~\micron, \OI~$1.129$~\micron\ / \FeI~$1.142$~\micron, Pa$\beta\,1.282$~\micron, and \SiI~$1.644$~\micron\ lines. The observations of SN\,2020jfo, SN\,2020qmp and SN\,2022acko are presented in Figure\,\ref{fig:SN_II_Neb_Line_Fits}. We have identified several strong emission features in the sample of nebular spectra and fit them with Lorentzian profiles, which fully capture the wings of the spectral features. The FWHM, and expansion velocity, of the line profile were determined from the profile fits. The H/He and O/Fe features displayed significant blending, especially the H/He emission feature. For the blended features, a combined double Lorentzian function was fitted, and bounds were set on the Lorentzian central peak location to ensure that the emission peaks were located close to the rest wavelength of the emission lines.

\begin{figure}
    \centering
    \includegraphics[width=\linewidth]{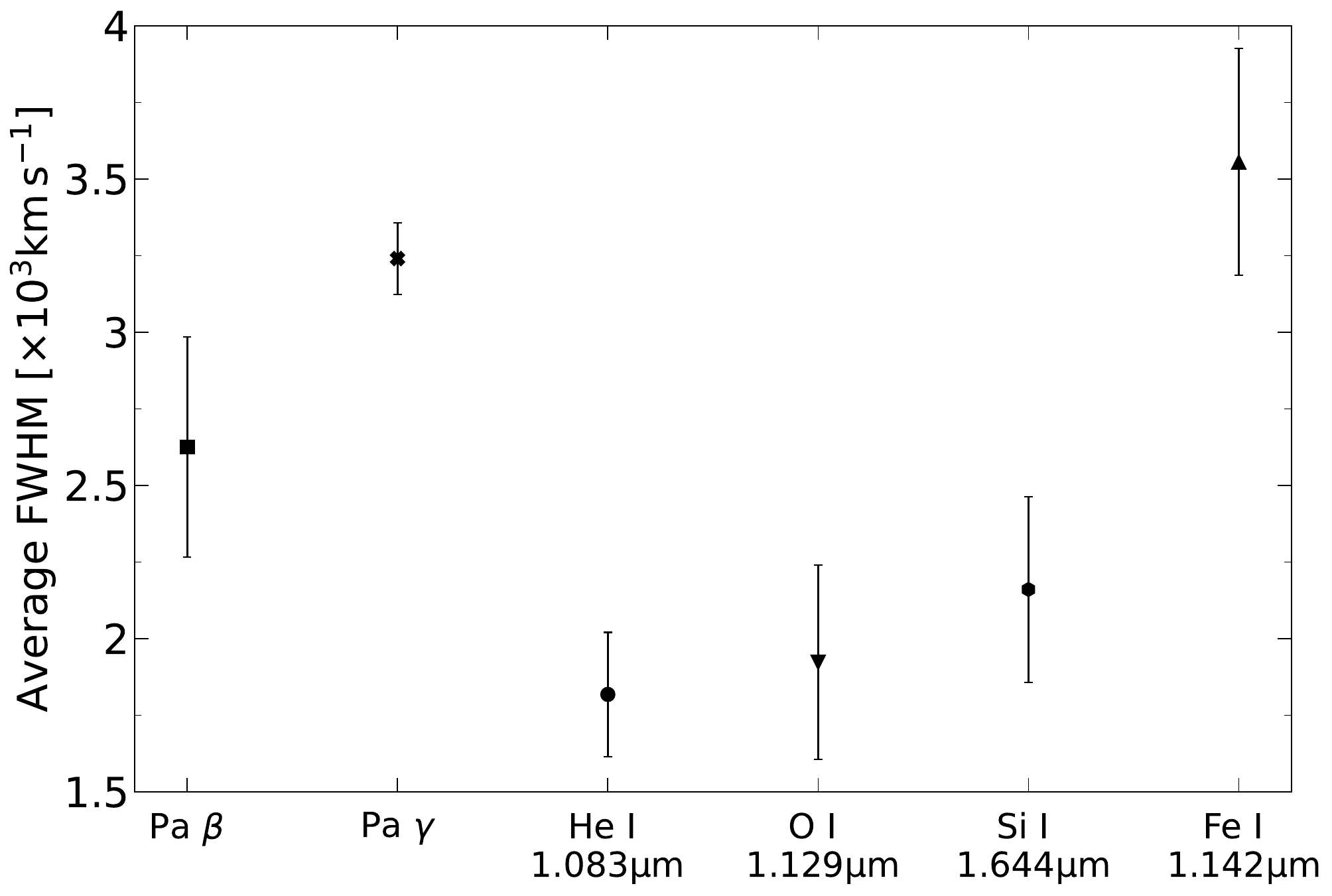}
    \caption{Average FWHM of the strong nebular phase emission lines across the 4 spectra of SNe\,2020jfo, 2020qmp, and 2022acko. The hydrogen envelope is located at larger velocities than all lines except the \FeI~$1.142$~\micron\ line, which are brought up significantly by SN\,2020jfo and SN\,2020qmp, suggesting a potential for mixing out of iron-group elements in these SNe, differences in metallicities, or asymmetries with the explosion.}
    \label{fig:FWHM_plots}
\end{figure}

The average FWHM's obtained from fitting the Lorentzian profiles to the nebular phase emission peaks are shown in Figure\,\ref{fig:FWHM_plots}. As expected, the hydrogen emission, with an average velocity of $2.6 \pm 0.4 \times10^3$~\kms\ and $3.2 \pm 0.1\times10^3$~\kms, originates from the outer regions of the ejecta at a higher velocity than the other emission features. The exception to this is the \FeI~$ 1.142$~\micron\ line, which has an average velocity of $3.6 \pm 0.4\times10^3$~\kms. The high velocity of the \FeI\ line of SN\,2020jfo and 2020qmp, likely arises from a blending out of iron. In contrast SN\,2022acko, a lower luminosity SN \citep{Bostroem_2023}, may have synthesized smaller amounts of \Nifs\ which is more tightly constrained to the inner slower core regions of the ejecta resulting in low \FeI\ line velocities. On top of this, differences in progenitor metallicities or asymmetries with the explosion may affect the chemical distribution in the ejecta. Line blending between the \OI~$1.129$~\micron\ and the \FeI~$1.142$~\micron\ lines may also influence the width of the \FeI\ line, causing an increase in the FWHM. Additional nebular phase NIR spectra of SNe\,II are required to determine the full diversity of how iron is distributed within the ejecta of SNe\,II, where a larger sample can also be used to understand the asymmetries within the explosion. 

\subsection{The First Overtone of CO in CC-SNe}

Prior to the formation of fresh dust within SN ejecta, molecular emission, most commonly from carbon monoxide (CO) and silicon monoxide (SiO), is detected through their fundamental and first overtone vibrational transitions \citep{Roche_1991, Gerardy_2002, Rho_2018, Shahbandeh_2024}. The CO first overtone, located between $2.3$ and $2.5$~\micron, is the only strong molecular feature in extragalactic SN spectra that is observable with ground-based instruments. This feature can be used to constrain the properties of the CO forming region, such as temperature, density, and CO mass \citep{Davis_2019, Mcleod_2024, Shahbandeh_2024, Stritzinger_2024}. Understanding the molecular chemistry in SN ejecta, particularly the production of CO, is crucial for tracing the formation of fresh dust and the broader process of dust enrichment in both the local and early universe. The epoch at which CO emerges in the spectra of SNe is dependent on how quickly the ejecta cools to below $\sim 2000$~K, as well as the structure of the envelope (i.e., when the carbon-rich core is visible). This typically occurs between $\sim +50 - 150$~days for CC-SNe. 

\begin{figure}
    \centering
    \includegraphics[width=\linewidth]{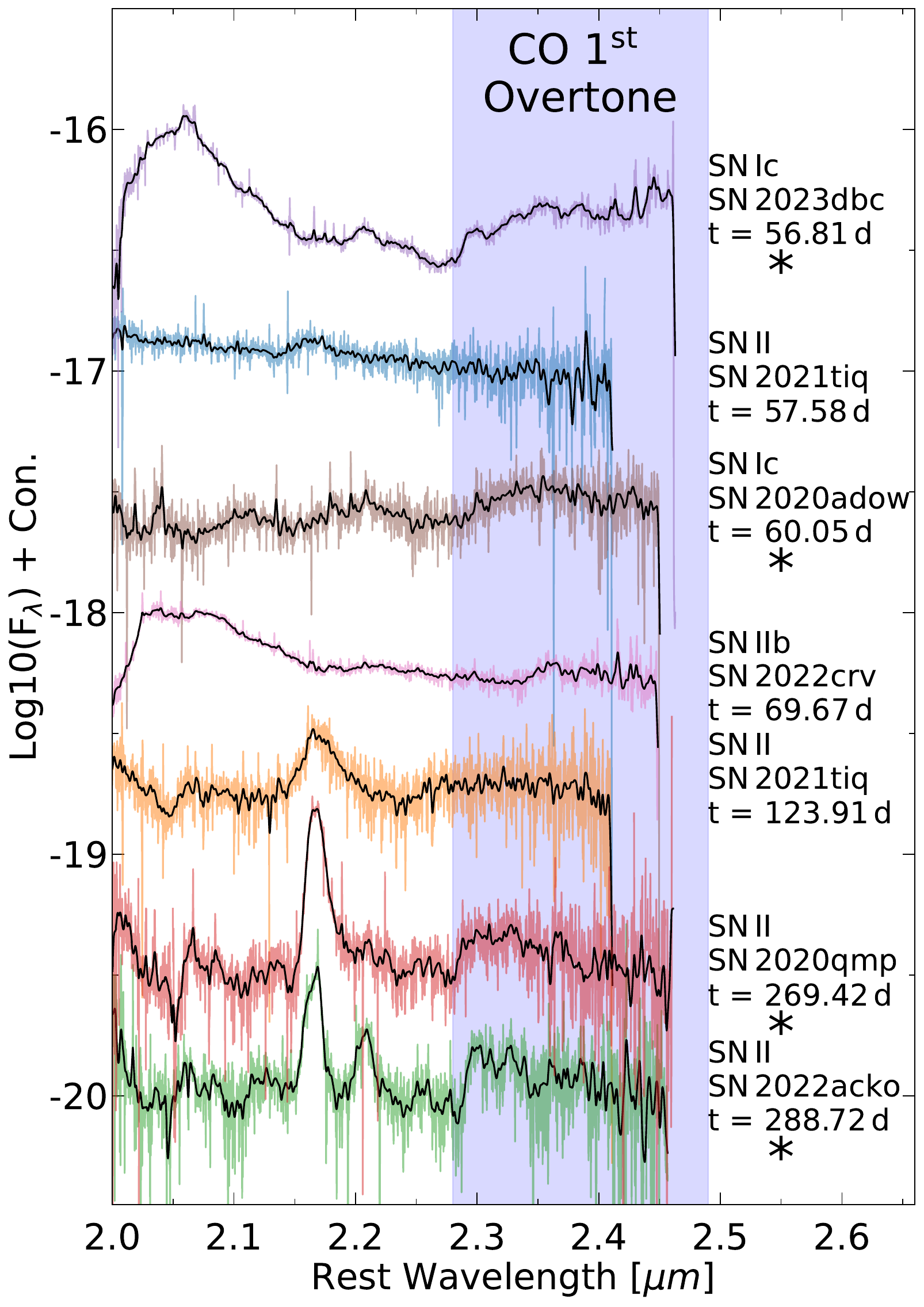}
    \caption{The CO first overtone can be identified in ground-based NIR spectra of CC-SNe in the weeks to months following maximum light. CO was not identified in all SNe shown; those SNe with confident identification of CO are marked with a $\mathrm{*}$.}
    \label{fig:CO_iden}
\end{figure}

Four of the SNe observed at phases greater than $50$~days from explosion possess features that are associated with the CO first overtone (SNe\,2020qmp, 2020adow, 2022acko, and 2023dbc), shown in Figure\,\ref{fig:CO_iden}. The emerging feature of the CO first overtone was observed in SNe\,2020adow and 2023dbc at $+60.05$ and $+56.81$~days respectively. CO is also observed significantly later in the spectra of SNe\,2020qmp and 2022acko, obtained at $+269.42$ and $+288.72$~days respectively suggesting significant amount of it has survived within the ejecta. 
Multiple observations of the CO first overtone allow for the density structure and temperature evolution of the ejecta to be determined more accurately constraining the mass of CO formed by the SN. 
Ground-based NIR observations of CC-SNe allow for the temporal coverage of the CO first overtone that is not possible with \textit{JWST}, due to the large time requirement and high cadence required to follow the evolution of the CO first overtone. The current literature data set of CO is sparse, and future observations, as well as modeling the current dataset,  can reveal the diversity in CO mass, its relationship to progenitor structure, and dust formation.

\begin{figure}
    \centering
    \includegraphics[width=\linewidth]{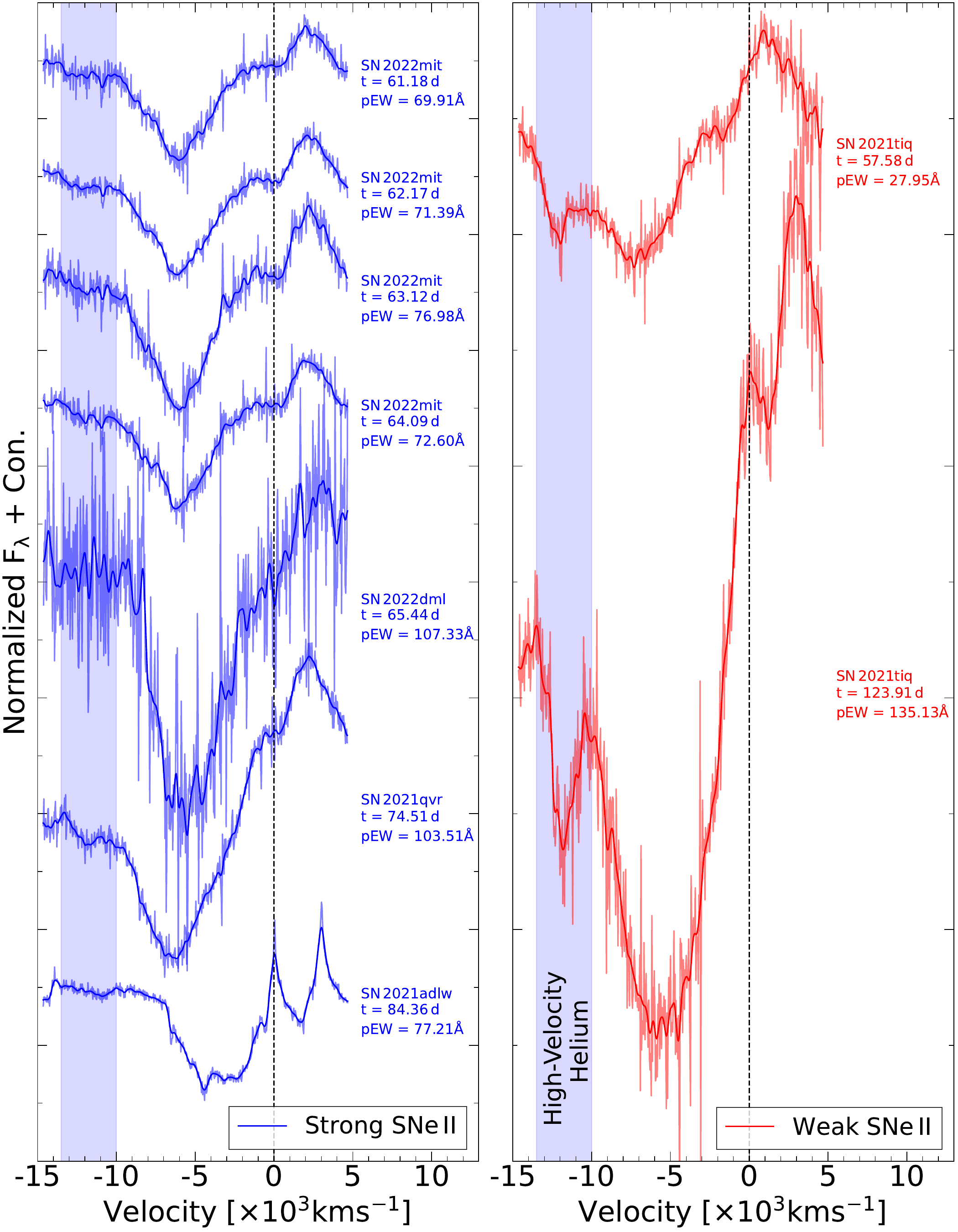}
    \caption{Line profile of the \HeI~$ 1.083$~\micron\ feature, used in determining the strength of the SN\,II. All spectra were taken between $+50 - 130$~days after maximum light, when the absorption component was still visible. Left: Strong SNe\,II SN\,2022mit, SN\,2022dml, SN\,2021qvr, and SN\,2021adlw, all of which possess a pEW $> 50$~\AA. These SNe lack a visible high-velocity component to the helium feature. Right: The weak SN\,II SN\,2021tiq.}
    \label{fig:Strong_vs_weak}
\end{figure}

\subsection{``Strong" vs ``Weak" SNe II}

The NIR spectra of SNe\,II typically display homogeneity in their features between explosion and roughly $+50$~days \citep{Davis_2019}. After this transition, they diverge into two distinct groups based on the pseudo-equivalent width (pEW) of the \HeI~$ 1.083$~\micron\ absorption feature.  ``Strong" SNe\,II, display pEWs $\mathrm{pEW} \geq 50$~\AA\ in the \HeI~$1.083$~\micron\ absorption feature, while SNe\,II with pEWs $\leq 50$~\AA\ are classified as ``weak". There are a number of differences between ``strong" and ``weak" SNe\,II \citep{Davis_2019}, the most visible of which is an absorption feature located on the blue side of the \HeI~$ 1.083$~\micron\ feature. This blue feature is associated with a high-velocity helium component. The dichotomy in SNe\,II was found to be linked to the structure of the progenitor and the photometric properties of the SN, where ``weak" SNe\,II decline in luminosity at much slower pace than ``strong" SNe\,II indicating the existence of two distinct groups of objects \citep{Davis_2019}. 

In the sample of SNe\,II presented in DR1, five events, SNe\,2022mit, 2022dml, 2021qvr, 2021adlw, and 2021tiq, were observed between $+50$ and $+130$~days, enabling measurement of the pEW and facilitating accurate sub-classification. For each SN the continuum surrounding the \HeI~$ 1.083$~\micron\ feature was fit and removed. The pEW was determined for the SNe\,II within the continuum-removed NIR spectra across the $\sim 1.0$~\micron\ region. These five SNe were sorted into ``weak" and ``strong", as shown in Figure\,\ref{fig:Strong_vs_weak}. SN\,2021tiq was the only SN\,II in our sample with a pEW less than $50$~\AA, while the rest all displayed much larger pEW. While the dataset for this region is still limited, it holds promise for expanding our understanding of progenitor structures and their CSM. 

\section{Conclusions} \label{sec:Summary}

In this work, we present the inaugural spectroscopic data release of \textit{HISS} of 48 transients. Our observations, acquired between 2021--2024 with NIRES on Keck-II and SpeX on IRTF, 
are comprised of 90 NIR spectra including: 30 SNe\,Ia, 33 SNe\,II, 14 SE-SNe, 8 interacting SNe, 4 spectra of the TDE AT\,2020ohl, and 1 spectrum of the SLSN-I SN\,2021fpl. We detail our data-reduction pipeline, identify the principal atomic and molecular transitions that are present in NIR spectra across different epochs and spectral types, and benchmark our sample against existing NIR templates to highlight its uniqueness.

We demonstrate how analysis of NIR spectra can advance four key science areas, which are:  
\textit{i)} The detection of unburnt carbon in SNe\,Ia,
\textit{ii)}  The measurement of the core structure of CC~SNe,
\textit{iii)}  The formation and evolution of CO in CC-SN ejecta, and
\textit{iv)}  The identification of high-velocity helium in CC~SNe.
This analysis can not be done with spectra at other wavelength regimes and highlights the importance of obtaining a large sample of thermonuclear and core-collapse SNe spectra in the NIR.

%What can be done with NIR spec
With the recent launch of \textit{JWST} and the forthcoming launch of \textit{Roman}, NIR astrophysics will remain at the forefront of transient research. However, space-based facilities alone cannot deliver the large spectroscopic samples needed to fully characterize these events; ground-based NIR follow-up is therefore indispensable.
In its next phase, \textit{HISS} will focus on the most exotic and poorly understood types of SNe, concentrating on the phases which are underrepresented in this data release, such as $t < -10$~days relative to maximum light, and at nebular phases. To accomplish this, \textit{HISS} will leverage the unparalleled capabilities of Maunakea NIR observing facilities including  
utilizing the available target-of-opportunity capabilities. Using these powerful tools, \textit{HISS} will continue  to advance a panchromatic understanding of cosmic explosions at the earliest and latest phases.

\section{Data Availability}

All spectra presented in this work are available online at the Weizmann Interactive Supernova Data Repository\footnote{\url{https://www.wiserep.org/}} \citep[WISeREP;][]{Yaron_2012}.

\section{Acknowledgments}

Some of the data presented herein were obtained at Keck Observatory, which is a private 501(c)3 non-profit organization operated as a scientific partnership among the California Institute of Technology, the University of California, and the National Aeronautics and Space Administration. The Observatory was made possible by the generous financial support of the W. M. Keck Foundation. 

We thank S.~Tinyanont for providing the custom \textsc{Python} scripts used in the flux calibration and telluric correction process of the \textsc{Pypeit} reduction.

K.M. and C.A. are supported by STScI grants (JWST-GO-02114, JWST-GO-02122, JWST-GO-04522, JWST-GO-03726, JWST-GO-6582, HST-AR-17555, JWST-GO-04217, JWST-GO-6023, JWST-GO-5290, JWST-GO-5057, JWST-GO-6677) and JPL-1717705. E.B., C.A., J.D., M.S., and  P.H. acknowledge support from NASA grants JWST-GO-02114, JWST-GO-02122, JWST-GO-04522, JWST-GO-04217, JWST-GO-04436, JWST-GO-03726, JWST-GO-05057, JWST-GO-05290, JWST-GO-06023, JWST-GO-06677, JWST-GO-06213, JWST-GO-06583. Support for programs \#2114, \#2122, \#3726, \#4217, \#4436, \#4522,  \#5057, \#6023, \#6213, \#6583, and \#6677 were provided by NASA through a grant from the Space Telescope Science Institute, which is operated by the Association of Universities for Research in Astronomy, Inc., under NASA contract NAS 5-03127. JH is funded by NASA grant 80NSSC23K1431.

% WBH NSF Acknowledgement
This material is based upon work supported by the National Science Foundation Graduate Research Fellowship Program under Grant Nos.\ 1842402 and 2236415. Any opinions, findings, conclusions, or recommendations expressed in this material are those of the authors and do not necessarily reflect the views of the National Science Foundation.

\vspace{5mm}
\facilities{Keck, IRTF}

\software{\textsc{astropy} \citep{Robitaille_2013,Price_Whelan_2018},  
          \textsc{pypeits} \citep{Prochaska_2020a, Prochaska_2020b},
          \textsc{Spextool} \citep{Cushing_2004}, and \textsc{xtellcor} \citep{Vacca_2003}
          }

%% Appendix material should be preceded with a single 
\appendix 

\section{Table of info}

Details of the SNe presented in DR1 and observed by \textit{HISS}, including right ascension, declination, classification, redshift, MJD date of maximum light, discovery date, number of observations, and discovery reference, are provided in the table below.
For the SNe that lacked ZTF $g$-band photometry, the source of the alternative data is indicated by a symbol next to the MJD value.
SNe that were reduced using the IDL-based \textsc{Spextool} are marked with a $^*$ symbol next to their name. These SNe were either observed with IRTF/SpeX or were too faint for reliable reduction with \textsc{Pypeit}.

\setcounter{table}{0}
\renewcommand{\thetable}{A\arabic{table}}

%\begin{rotatetable}
\startlongtable
\begin{deluxetable}{ccccccccc}%{rlllllll}
%\tablewidth{0pt}
\tablecaption{Details on \textit{HISS} transients observed between 2021-2024. Unless indicated the $t_{max}$ refers to the MJD date of $g$-band maximum used to determine the phase of observation. 
When ZTF data was not available additional sources were used to determine $t_{max}$: 
$a$ = ATLAS $c$-band used,
$b$ = sourced from \citet{Hinkle_2022},
$c$ = No peak photometry, discovery date used instead,
$d$ = sourced from \citet{Zhao_2024},
$e$ = ZTF $r$-band used,
$f$ = sourced from \citet{Zimmerman_2024}.
$^*$ Spectra reduced by IDL-based \textsc{Spextool}.
}
\label{tab:SN_info}
\tablehead{
\colhead{SN} & \colhead{Ra (J2000)} & \colhead{Dec (J2000)} & \colhead{Type} & \colhead{Redshift} & \colhead{$t_{max}$} & \colhead{Disc. MJD} & \colhead{No. Obs} & \colhead{\hspace{-.5cm}Discovery Report}
}
\startdata
    2020jfo & 12:21:50.480 & +04:28:54.05 & II & 0.0050 & 58983.62 & 59310.24 & 1 & \cite{Nordin_2020} \\
    \multirow{2}{*}{ASASSN-20hx/} & \multirow{2}{*}{17:03:36.492} & \multirow{2}{*}{+62:01:32.34} & \multirow{2}{*}{ANT/TDE} & \multirow{2}{*}{0.0167} & \multirow{2}{*}{59052.80$^b$} & \multirow{2}{*}{59040.34} & \multirow{2}{*}{4} & \multirow{2}{*}{\cite{Stanek_2020a}} \\
    2020ohl$^*$ & & & & & & & & \\
    2020qmp & 12:08:44.440 & +36:48:17.28 & II & 0.0030 & 59366.34$^c$ & 59366.34 & 1 & \cite{De_2020} \\
    2020aatb & 05:24:46.412 & $-$15:56:31.43 & II & 0.0100 & 59192.52 & 59176.37 & 3 & \cite{Munoz-Arancibia_2020} \\
    2020acat & 11:50:41.225 & $-$10:13:38.92 & IIb & 0.0079 & 59206.67 & 59192.63 & 1 & \cite{Tonry_2020b} \\
    2020adow & 08:33:42.262 & +27:42:43.56 & Ic & 0.0075 & 59216.74 & 59209.41 & 2 & \cite{Stanek_2020b} \\
    2020aekp & 15:43:11.390 & +17:48:47.20 & Ia-CSM & 0.046 & 59223.83 & 59529.50 & 1 & \cite{Hodgkin_2020} \\
    2021csp & 14:26:22.114 & +05:51:33.18 & Icn & 0.0830 & 59258.90 & 59256.51 & 1 & \cite{Perley_2021} \\
    2021dbg & 09:24:26.790 & $-$06:34:53.15 & II & 0.0200 & 59269.32$^d$ & 59260.43 & 1 & \cite{Tonry_2021a} \\
    2021dov$^*$ & 08:56:18.710 & $-$00:26:31.70 & Ia & 0.0120 & 59283.87 & 59267.28 & 3 & \cite{Fremling_2021} \\
    2021foa$^*$ & 07:57:03.256 & +46:27:45.83 & IIn & 0.0084 & 59302.00$^a$ & 59288.45 & 1 & \cite{Forster_2020} \\
    2021fpl & 20:14:18.620 & $-$18:10:56.57 & SLSN-I & 0.1150 & 59334.48 & 59288.63 & 1 & \cite{Tonry_2021b} \\
    2021gno$^*$ & 12:12:10.290 & +13:14:57.05 & Ib & 0.0062 & 59305.51 & 59293.24 & 2 & \cite{Bruch_2021} \\
    2021hem$^*$ & 16:21:16.003 & +14:33:09.65 & Ia & 0.0350 & 59316.41 & 59297.42 & 1 & \cite{Forster_2021a} \\
    2021hiz$^*$ & 12:25:41.670 & +07:13:42.20 & Ia & 0.0033 & 59320.37 & 59303.36 & 1 & \cite{Munoz-Arancibia_2021a} \\
    2021kat & 19:50:31.660 & +57:59:28.07 & IIn & 0.1013 & 59360.22 & 59317.49 & 1 & \cite{Forster_2021b} \\
    2021koq$^*$ & 18:16:01.020 & +06:45:06.37 & Ia & 0.0205 & 59345.55 & 59332.48 & 1 & \citep{Tonry_2021j} \\
    2021mxs & 12:55:26.192 & +52:15:55.37 & Ia & 0.0344 & 59369.88 & 59355.26 & 1 & \cite{Chambers_2021} \\
    2021pfs$^*$ & 14:03:23.580 & $-$06:01:53.90 & Ia & 0.0087 & 59391.81 & 59374.22 & 2 & \cite{Munoz-Arancibia_2021b} \\
    2021pqk$^*$ & 14:19:15.580 & +26:18:01.40 & Ia & 0.0368 & 59392.63 & 59394.39 & 1 & \cite{De_2021b} \\
    2021qvr$^*$ & 23:28:26.650 & +22:25:11.89 & II & 0.0116 & 59399.00$^a$ & 59387.59 & 2 & \cite{Tonry_2021d} \\
    2021qvv & 12:28:02.920 & +09:48:10.26 & Ia-91bg-like & 0.0018 & 59395.31 & 59388.26 & 2 & \cite{Tonry_2021e} \\
    2021rhu & 00:03:15.429 & +16:08:44.55 & Ia & 0.0035 & 59411.48 & 59396.41 & 1 & \cite{Munoz-Arancibia_2021c} \\
    2021tiq & 22:36:54.720 & $-$12:33:41.98 & II & 0.0239 & 59415.39 & 59409.42 & 3 & \cite{Munoz-Arancibia_2021d} \\
    2021uqw & 21:53:59.427 & +06:41:51.06 & IIb & 0.0280 & 59431.02 & 59428.32 & 1 & \cite{Munoz-Arancibia_2021d} \\
    2021yja & 03:24:21.180 & $-$21:33:56.20 & IIP & 0.0053 & 59475.44 & 59465.55 & 1 & \cite{Tonry_2021f} \\
    2021acnt & 09:41:42.840 & $-$28:12:08.96 & Ia & 0.0080 & 59529.60$^a$ & 59513.62 & 3 & \cite{Tonry_2021g} \\
    2021adlr & 03:11:11.760 & $-$08:28:10.42 & II & 0.0296 & 59537.01 & 59522.45 & 1 & \cite{Tonry_2021h} \\
    2021adlv & 08:31:03.830 & $-$04:11:39.19 & Ia & 0.0330 & 59535.48 & 59523.48 & 1 & \cite{Nordin_2021} \\
    2021adlw & 11:41:17.822 & +36:32:34.57 & II & 0.0049 & 59544.86 & 59523.51 & 1 & \cite{Munoz-Arancibia_2021f} \\
    2021aess & 03:33:48.140 & $-$19:29:44.52 & IIn-pec & 0.0070 & 59544.00$^a$ & 59532.27 & 1 & \cite{De_2021} \\
    2021agef & 06:12:31.278 & +06:41:52.01 & Ia & 0.0130 & 59567.00$^a$ & 59557.42 & 2 & \cite{Tonry_2021i} \\
    2022crv & 09:54:25.910 & $-$25:42:11.16 & IIb & 0.0080 & 59654.00$^a$ & 59629.19 & 3 & \cite{Dong_2022} \\
    2022dml & 16:17:29.120 & +14:25:04.62 & II & 0.0300 & 59648.20 & 59637.45 & 1 & \cite{Fremling_2022a} \\
    2022fcc & 14:15:54.759 & +03:36:14.60 & Ia & 0.0259 & 59680.00$^a$ & 59663.26 & 1 & \cite{Tonry_2022b} \\
    2022hrs & 12:43:34.335 & +11:34:35.87 & Ia-HV & 0.0047 & 59699.88 & 59303.20 & 1 & \cite{Itagaki_2022} \\
    2022ihz & 09:42:48.220 & $-$03:36:25.45 & Ia-91bg-like & 0.0060 & 59705.72 & 59694.31 & 1 & \cite{Tonry_2022c} \\
    2022jli & 00:34:45.690 & $-$08:23:12.16 & Ic & 0.0060 & 59752.00$^a$ & 59704.17 & 3 & \cite{Monard_2022} \\
    2022jzx & 01:02:09.860 & $-$04:14:06.61 & II & 0.0170 & 59770.13 & 59715.41 & 4 & \cite{Tonry_2022d} \\
    2022mbg & 02:06:02.990 & +29:47:31.13 & Ia & 0.0162 & 59738$^e$ & 59734.47 & 4 & \cite{Fremling_2022b} \\
    2022mit & 20:58:55.410 & $-$27:25:12.54 & II & 0.0200 & 59748.00$^a$ & 59738.57 & 4 & \cite{Chambers_2022} \\
    2022mww & 03:12:47.060 & +41:49:13.19 & Ia & 0.0200 & 59757.47 & 59747.47 & 4 & \cite{Munoz-Arancibia_2022a} \\
    2022ojo & 01:44:35.620 & +37:41:50.68 & IIP & 0.0190 & 59781.41 & 59765.57 & 4 & \cite{Tonry_2022e} \\
    2022prr & 19:01:41.897 & +40:45:03.67 & IIn & 0.0152 & 59794.42 & 59787.30 & 3 & \cite{Stanek_2022} \\
    2022pux & 21:21:07.850 & +23:04:58.51 & II & 0.0157 & 59796.20 & 59788.39 & 4 & \cite{Munoz-Arancibia_2022b} \\
    2022acko & 03:19:38.990 & $-$19:23:42.68 & II & 0.0053 & 59922.23 & 59919.16 & 2 & \cite{Lundquist_2022} \\
    2023dbc & 11:11:39.230 & +55:40:28.99 & Ic & 0.0023 & 60034.35 & 60016.29 & 1 & \cite{Ho_2023} \\
    2023wne & 04:39:57.919 & $-$08:55:11.51 & Ic & 0.0170 & 60264.46 & 60251.42 & 1 & \cite{De_2023} \\
\enddata
%\tablecomments{Details on HISS transients.}
\end{deluxetable}
%\end{rotatetable}

\clearpage

\section{Figures}

The figures below show the spectra of the transients observed in \textit{HISS} DR1. The supernovae have been grouped into their spectral classifications: SNe\,Ia, SN\,II, SE-SNe, interacting SNe, TDE, and SLSN-I. Additionally, the SNe\,II have been separated into early-phase spectra ($t < 150$ days; Figure\,\ref{fig:II_phot}) and late-phase spectra ($t > 150$ days; Figure\,\ref{fig:II_neb}) to highlight the different physical processes that dominate each phase. All spectra have been corrected for Milky Way extinction and are presented in the rest frame. A one-dimensional Gaussian filter has been applied to smooth each spectrum, shown by the solid black line overlaid on the raw flux data, which is plotted in gray. Phases are given relative to $t_{max}$, as listed in Table~\ref{tab:SN_info}. The two near-infrared telluric absorption regions have been masked with gray vertical bars to enhance the clarity of the plots.

\setcounter{figure}{0}
\renewcommand{\thefigure}{B\arabic{figure}}
\renewcommand{\theHfigure}{B{\arabic{figure}}}

\begin{figure}[!h]
    \centering
    \includegraphics[width=0.9\linewidth%, height=\textheight
    ]{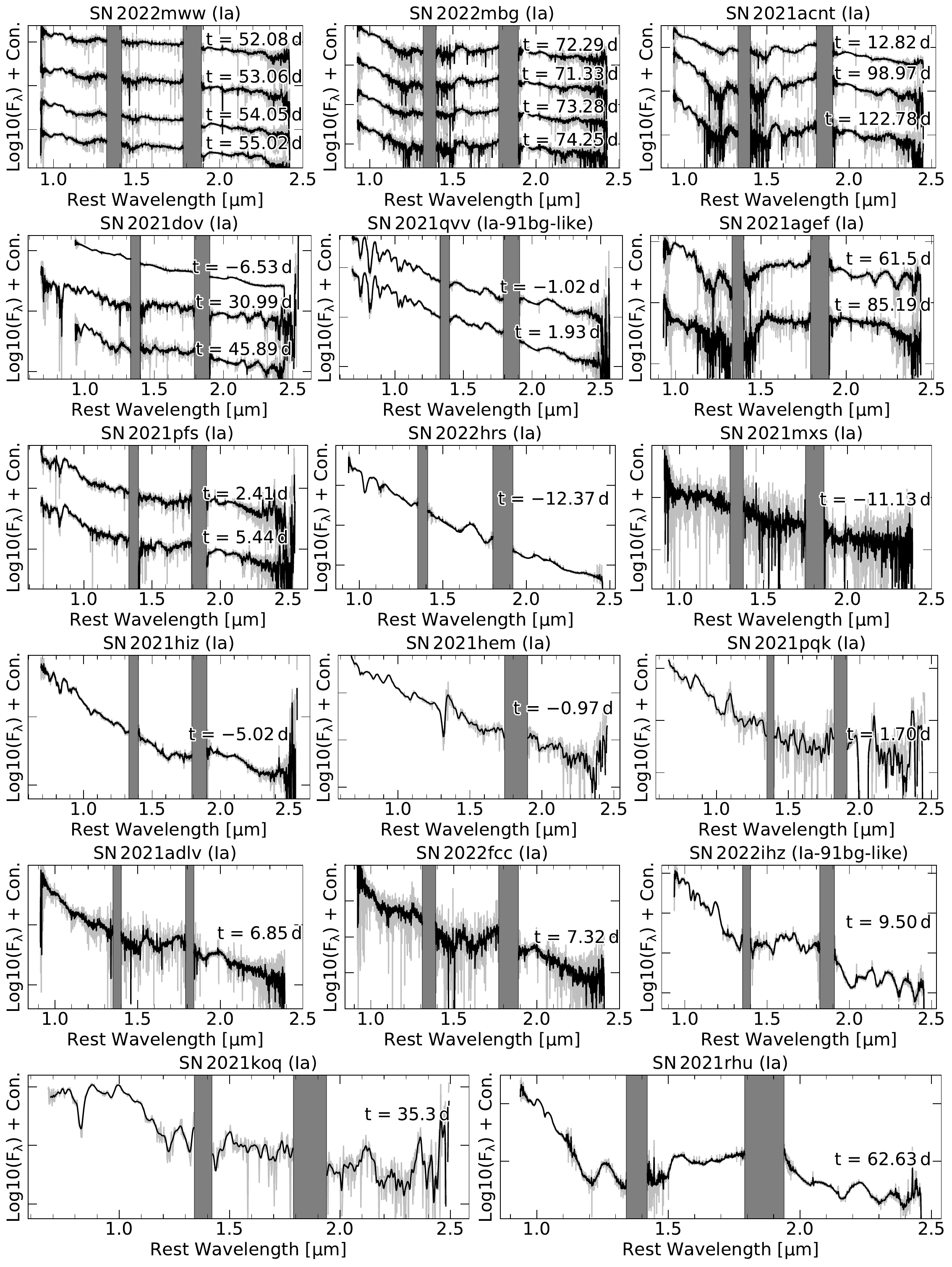}
    \caption{\textit{HISS} spectra of SNe\,Ia ordered by number of observations and phase, with the spectral phase shown above each spectrum. All spectra are shown in the rest frame, and the spectral classification is given for each object. Spectra have been smoothed using a Gaussian filter with $\sigma=3$, shown by the solid black line, with the raw spectra shown by the gray line. Telluric regions are shown by the gray vertical regions.}
    \label{fig:Ia_phot}
\end{figure}

\begin{figure*}
    \centering
    \includegraphics[width=0.95\linewidth]{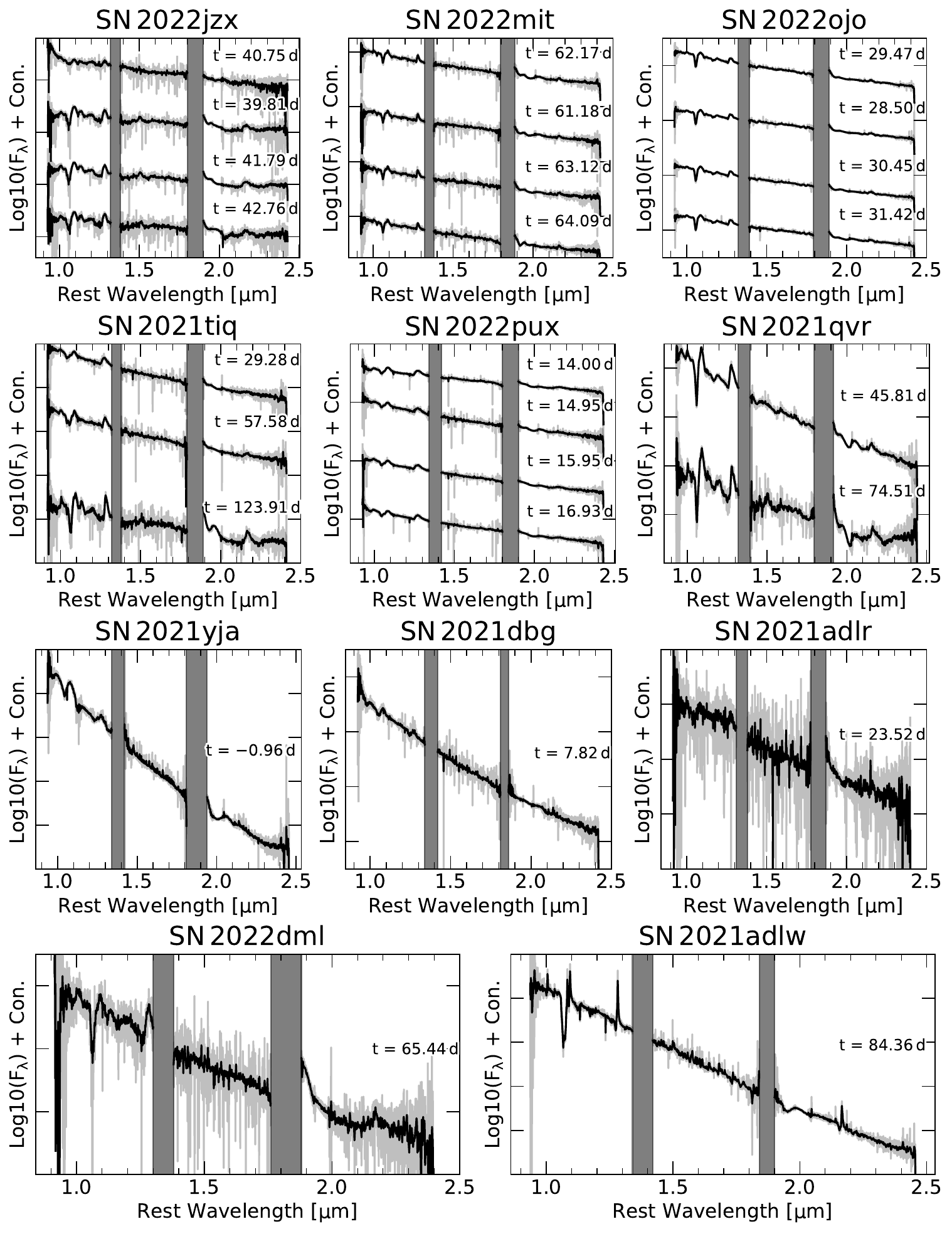}
    \caption{Same as Figure\,\ref{fig:Ia_phot} but for SNe\,II observed before $+150$~days after maximum light.}
    \label{fig:II_phot}
\end{figure*}

\begin{figure*}
    \centering
    \includegraphics[width=0.95\linewidth]{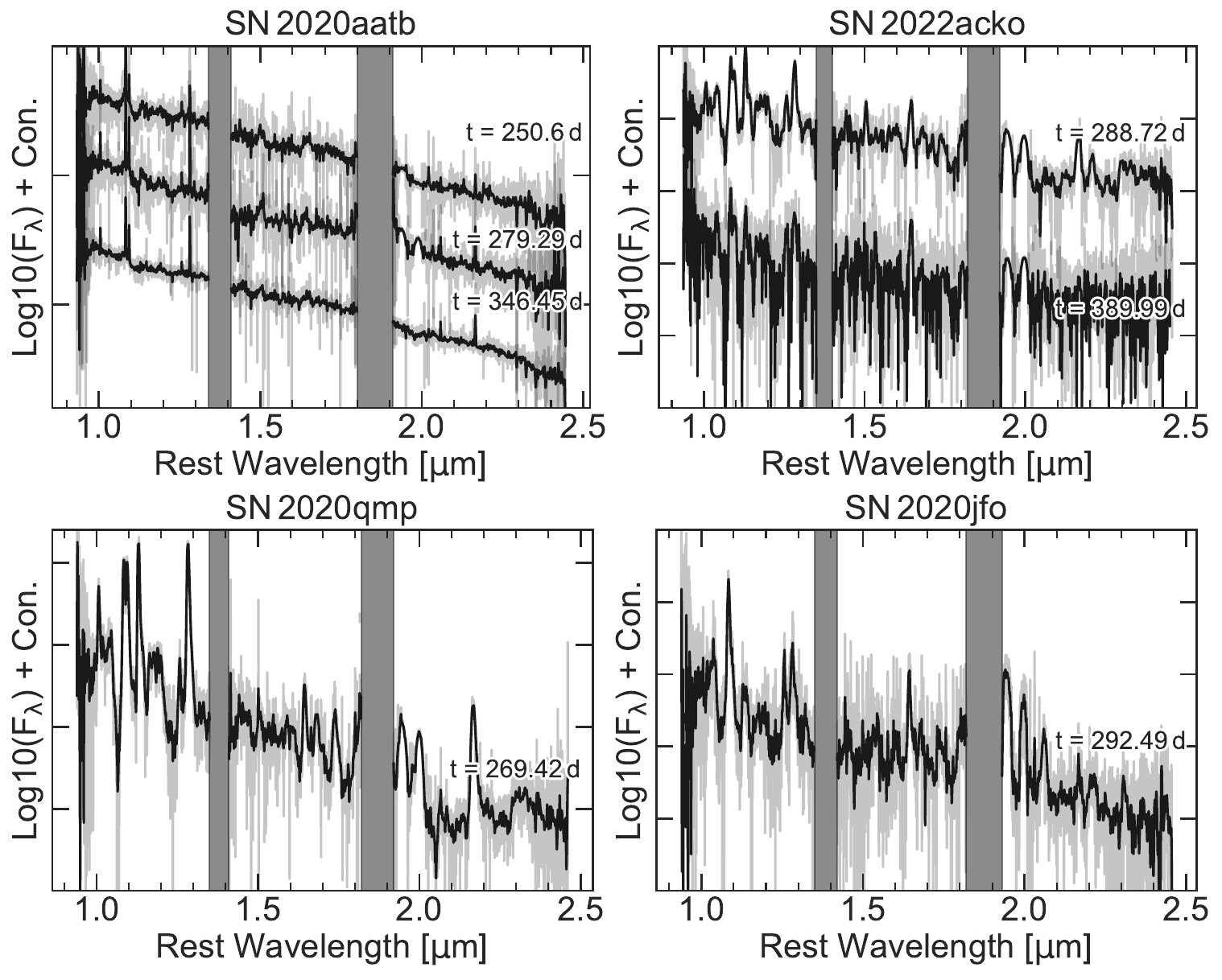}
    \caption{Same as Figure\,\ref{fig:Ia_phot} but for SNe\,II observed at $t \geq+150$~days after maximum light.}
    \label{fig:II_neb}
\end{figure*}

\begin{figure*}
    \centering
    \includegraphics[width=0.95\linewidth, height=22cm]{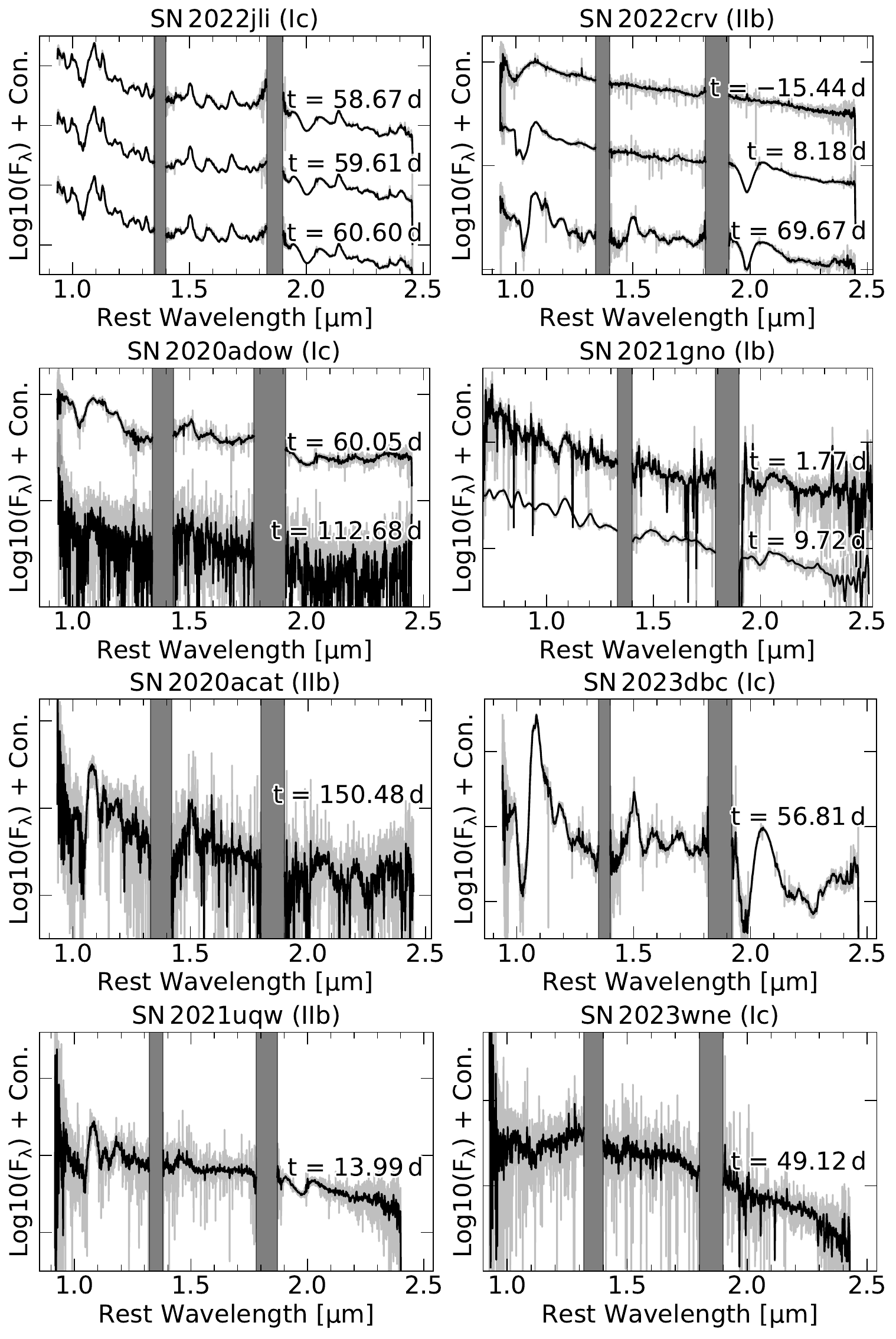}
    \caption{Same as Figure\,\ref{fig:Ia_phot} but for SE-SNe.}
    \label{fig:SE-SNe}
\end{figure*}

\begin{figure*}
    \centering
    \includegraphics[width=0.95\linewidth]{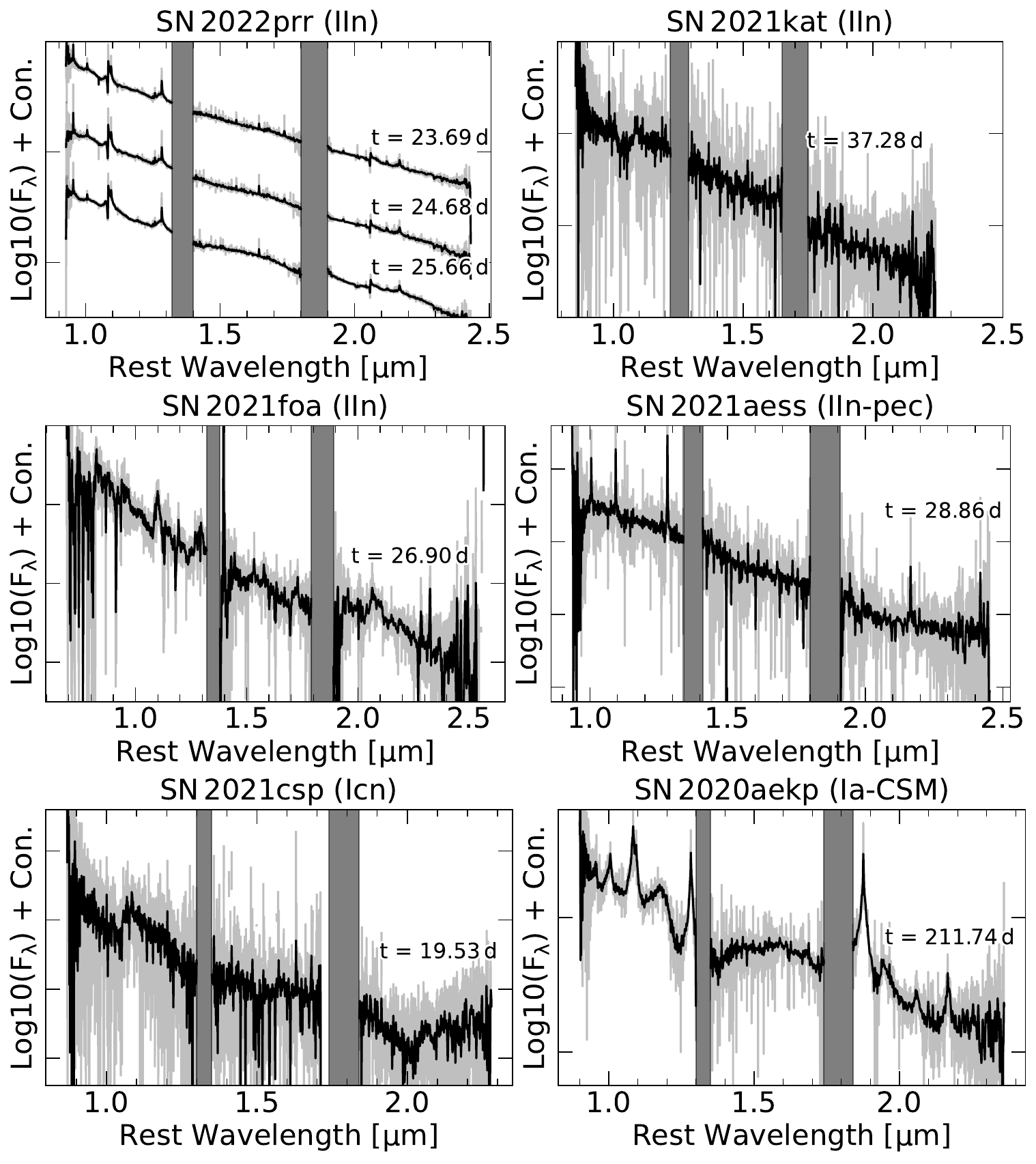}
    \caption{Same as Figure\,\ref{fig:Ia_phot} but for Interacting SNe, including SNe\,Ia-CSM, SNe\,IIn, and the SN\,Icn SN\,2021csp.}
    \label{fig:Interacting}
\end{figure*}

\begin{figure*}
    \centering
    \includegraphics[width=0.95\linewidth]{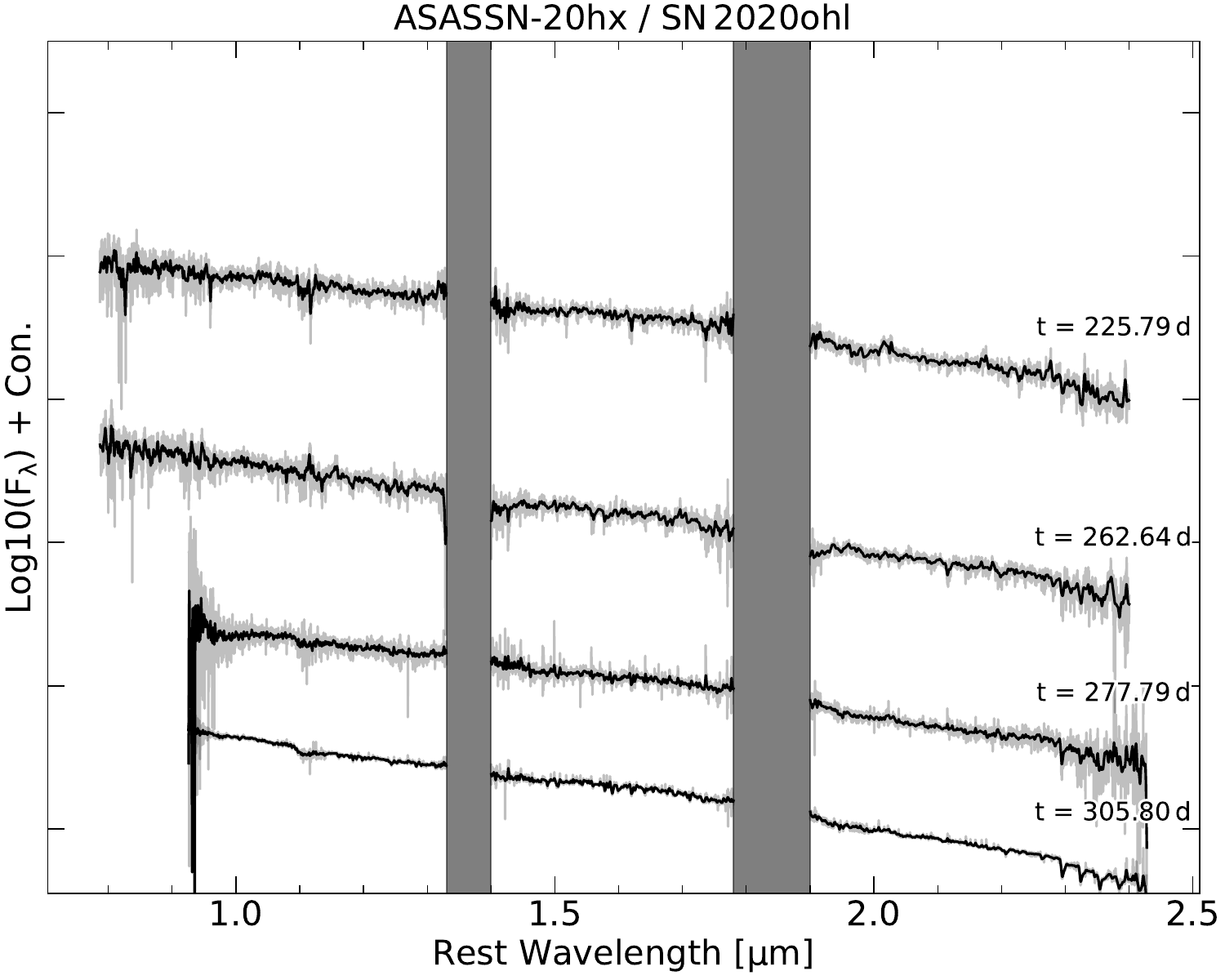}
    \caption{Same as Figure\,\ref{fig:Ia_phot} but for the TDE ASASSN-20hx/AT2020ohl. The spectra are blue and featureless throughout the evolution of ASASSN-20hx.}
    \label{fig:TDE_spec}
\end{figure*}

\begin{figure*}
    \centering
    \includegraphics[width=0.95\linewidth]{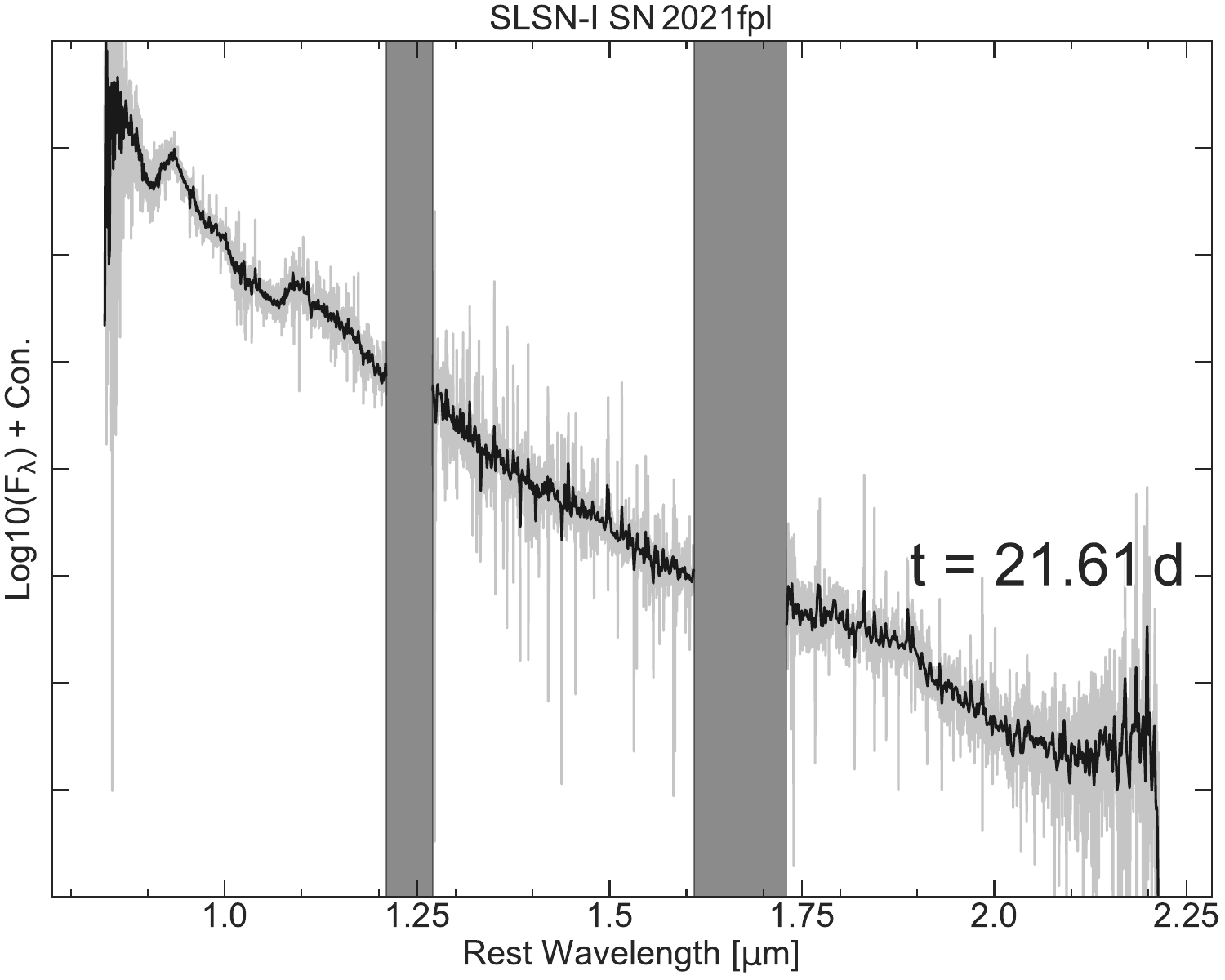}
        \caption{Same as Figure\,\ref{fig:Ia_phot} but for the SLSN-I SN\,2020fpl.}
    \label{fig:SLSN_spec}
\end{figure*}

\newpage

\bibliography{Main}{}
\bibliographystyle{aasjournal}

\end{document}